\documentclass[review]{elsarticle}
\usepackage{graphicx}
\begin{document}

\begin{frontmatter}
\title{Effect of Small-World Connectivity on Fast Sparsely Synchronized Cortical Rhythms}
\author{Sang-Yoon Kim}
\ead{sangyoonkim@dnue.ac.kr}
\author{Woochang Lim\corref{mycorrespondingauthor}}
\cortext[mycorrespondingauthor]{Corresponding author}
\ead{woochanglim@dnue.ac.kr}
\address{Computational Neuroscience Lab., Department of Science Education, Daegu National University of Education, Daegu 705-115, Korea}

\begin{abstract}
Fast cortical rhythms with stochastic and intermittent neural discharges have been observed in electric recordings of brain activity. For these fast sparsely synchronized oscillations, individual neurons fire spikings irregularly and sparsely as Geiger counters, in contrast to fully synchronized oscillations where individual neurons exhibit regular firings like clocks. We study the effect of network architecture on these fast sparsely synchronized rhythms in an inhibitory population of suprathreshold fast spiking (FS) Izhikevich interneurons (which fire spontaneously without noise). We first employ the conventional Erd\"{o}s-Renyi random graph of suprathreshold FS Izhikevich interneurons for modeling the complex connectivity in neural systems, and study emergence of the population synchronized states by varying both the synaptic inhibition strength $J$ and the noise intensity $D$. Fast sparsely synchronized states of relatively high degree are found to appear for large values of $J$ and $D$. However, in a real cortical circuit, synaptic connections are known to have complex topology which is neither regular nor random. Hence, for fixed values of $J$ and $D$ we consider the Watts-Strogatz small-world network of suprathreshold FS Izhikevich interneurons which interpolates between regular lattice and random graph via rewiring, and investigate the effect of small-world synaptic connectivity on emergence of fast sparsely synchronized rhythms by varying the rewiring probability $p$ from short-range to long-range connection. When passing a small critical value $p^*_c$, fast sparsely synchronized population rhythms are found to emerge in small-world networks with predominantly local connections and rare long-range connections. This transition to fast sparse synchronization is well characterized in terms of a realistic ``thermodynamic'' order parameter. For further understanding of this transition, we also investigate the effect of long-range connections on dynamical correlations between neuronal pairs, and find that for $p>p^*_c$, global synchronization appears in the whole population because the spatial correlation length covers the whole system thanks to sufficient number of long-rang connections. The degree of fast sparse synchronization for $p > p^*_c$ is also measured in terms of a realistic ``statistical-mechanical'' spiking measure. As $p$ is increased from $p^*_c$, the degree of population synchrony becomes higher, while the axon ``wire length'' of the network also increases. At a dynamical-efficiency optimal value $p_{\cal{E}}^*$, there is a trade-off between the population synchronization and the wiring economy, and hence an optimal fast sparsely-synchronized rhythm is found to occur at a minimal wiring cost in an economic small-world network.
\end{abstract}

%\pacs{87.19.lm, 87.19.lc}
\begin{keyword}
Suprathreshold FS Izhikevich interneurons \sep Small-world network \sep Fast sparsely synchronized cortical rhythm
\end{keyword}

\end{frontmatter}

\section{Introduction}
\label{sec:INT}
Recently, brain rhythms have attracted  much attention \citep{Buz1}. Particularly, we are interested in fast sparsely synchronized cortical rhythms, associated with diverse cognitive functions \cite{W_Review,TW}. In some experimental data \cite{SS1,SS2,SS3,SS4,SS5,SS6,SS7},
synchronous small-amplitude fast oscillations [e.g., gamma rhythm (30-100 Hz) during awake behaving states and rapid eye movement sleep and
sharp-wave ripple (100-200 Hz) during quiet sleep and awake immobility] have been observed in local field potential recordings, while individual neuron recordings have been found to show stochastic and intermittent spike discharges. Thus, single-cell firing activity differs markedly from the population oscillatory behavior. We note that these sparsely synchronized rhythms are in contrast to fully synchronized rhythms. For the fully synchronized rhythms, individual neurons fire regularly at the population frequency like the clock oscillators \cite{WB}. Hence, fully synchronized oscillations may be well described by using the conventional coupled-oscillator model composed of suprathreshold spiking neurons above a threshold in the absence of noise or for weak noise \cite{Wang}. However, such coupled-oscillator models seem to be inappropriate for describing sparse synchronization because of stochastic and intermittent individual neural discharges like the Geiger counters. By taking an opposite view from that of coupled oscillators, the authors in \cite{Sparse1,Sparse2,Sparse3,Sparse4,Sparse5,Sparse6,TJ1,TJ2} developed a framework appropriate for description of sparse synchronization. When the external noise is strong, suprathreshold spiking neurons discharge irregular firings as Geiger counters, and then the population state becomes unsynchronized. However, as the inhibitory recurrent feedback becomes sufficiently strong, a synchronized population state with stochastic and sparse neural discharges emerges. In this way, under the balance between strong external excitation and strong recurrent inhibition, fast sparse synchronization was found to occur in networks of suprathreshold neurons for both cases of random coupling \cite{Sparse1,Sparse2,Sparse3,Sparse4} and global coupling \cite{Sparse5,Sparse6,TJ1,TJ2}. Similar sparsely synchronized rhythms were also found to appear through via cooperation of noise-induced spikings of subthreshold Morris-Lecar neurons (which can not fire spontaneously without noise) \cite{Kim1,Kim2,Kim3}.

In this paper, we study the effect of network architecture on fast sparsely synchronized cortical rhythms in an inhibitory population of suprathreshold fast spiking (FS) Izhikevich interneurons \cite{Izhi1,Izhi2,Izhi3,Izhi4}. The conventional Erd\"{o}s-Renyi random graph has been
usually used for modeling complex connectivity occurring in diverse fields such as social, biological, and technological networks \cite{ER}.
So, we first consider a random network of suprathreshold FS Izhikevich interneurons, and investigate occurrence of the population synchronized states by varying the inhibition strength and the noise intensity. Fast sparsely synchronized oscillations are found to appear when both the inhibition and the noise are sufficiently strong. Global efficiency of information transfer becomes high for random connection because its average path length (i.e., typical separation between two neurons represented by average number of synapses between two neurons along the minimal path) is short due to long-range connections \cite{Eff1,Eff2}. However, random networks have poor clustering (i.e., low cliquishness of a typical neighborhood) and they are non-economic ones because the (axon) wiring cost becomes expensive due to appearance of short-range and long-range connections with equal probability \cite{Sporns,Buz2}. In a real cortical circuit, synaptic connections are known to have complex topology which is neither regular nor completely random \cite{Sporns,Buz2,CN1,CN2,CN3,CN4,CN5,CN6,CN7}. Hence, we consider the Watts-Strogatz model for small-world networks which interpolates between regular lattice with high clustering and random graph with short path length via rewiring \cite{SWN1,SWN2,SWN3}. The Watts-Strogatz model may be regarded as a cluster-friendly extension of the random network by reconciling the six degrees of separation (small-worldness) \cite{SDS1,SDS2} with the circle of friends (clustering). Many recent works on various subjects of neurodynamics have been done in small-world networks with predominantly local connections and rare long-distance connections \cite{CN6,SW2,SW3,SW4,SW5,SW6,SW7,SW8,SW9,SW10,SW11,SW12,SW13}. Here, we investigate the effect of small-world connectivity on emergence of fast sparsely synchronized rhythms by varying the rewiring probability $p$ from local to long-range connections. As $p$ is increased, long-range short-cuts that connect distant neurons begin to appear, and the average path length can be dramatically decreased only by a few short-cuts. Thus, global effective communication between distant neurons may be available via shorter synaptic paths. Eventually, when $p$ passes a critical value $p_c^*$, fast sparsely synchronized rhythm emerges in the whole population because dynamical correlation length covers the whole system thanks to sufficient number of long-rang connections. However, with increasing $p$, the (axon) wiring length also becomes longer due to appearance of long-range connections. Longer axonal projections are expensive due to their material and energy costs. Hence, we must take into account the (axon) wiring economy for the dynamical efficiency because wiring cost is an important constraint of the brain evolution \cite{Buz1,W_Review,Sporns,Kim3,Buz2,SW4,SW7,Wiring1,Wiring2,Wiring3,Wiring4,Wiring6,Wiring7}. At a dynamical-efficiency optimal value $p_{\cal{E}}^*$ an optimal fast sparse synchronization is found to occur via trade-off between synchrony and wiring cost at a minimal wiring cost in an economic small-world network \cite{Buz2}.

This paper is organized as follows. In Sec.~\ref{sec:SWN}, we describe an inhibitory population of suprathreshold FS Izhikevich interneurons. The Izhikevich neurons are not only biologically plausible, but also computationally efficient \cite{Izhi1,Izhi2,Izhi3,Izhi4}, and they interact through inhibitory GABAergic synapses (involving the $\rm {GABA_A}$ receptors). In Sec.~\ref{sec:SSR}, we first consider the conventional Erd\"{o}s-Renyi random graph \cite{ER}, and study appearance of the population synchronized states by varying the noise intensity $D$ and the inhibition strength $J$. We fix $J$ and $D$ at appropriately strong values where sparsely synchronized rhythms of relatively high degree emerge. Then, we consider the Watts-Strogatz model for the small-world network which interpolates between the regular lattice and the random graph \cite{SWN1}, and investigate the effect of the small-world connectivity on fast sparsely synchronized rhythms by increasing the rewiring probability $p$. For the regular connection of $p=0$, the average path length is very long because there exist only short-range connections, and hence an unsynchronized population state appears. However, with increasing $p$, long-range connections begin to appear, and hence the average path length becomes shorter. Consequently, when passing a critical value $p_c^*$ $(\simeq 0.12)$, the unsynchronized state is destabilized and then fast sparsely synchronized population rhythm emerges. This transition to fast sparse synchronization is well described by using a realistic ``thermodynamic'' order parameter, based on the instantaneous population spike rate (IPSR) \cite{RM}. In order to further investigate the effect of geometrical long-range connections on dynamical spatial correlation length for occurrence of population synchronization, we also make a dynamical-correlation analysis between neuronal pairs. It is thus found that the spatial correlation length for $p < p_c^*$ is so small that global synchronization cannot occur. On the other hand, for $p > p_c^*$, the correlation length is found to cover the whole system thanks to sufficient number of long-rang connections, and consequently global synchronization appears in the whole population. Furthermore, the degree of fast sparse synchronization is well measured by employing a realistic ``statistical-mechanical'' spiking measure, based on IPSR \cite{RM}. At an optimal value $p^*_{\cal{E}}$ $(\simeq 0.26)$, a dynamical-efficiency factor given by the ratio of the synchronization degree to the (axon) wiring cost is found to become maximal.  Thus, an optimal fast sparsely-synchronized rhythm is found to appear at a minimal wiring cost in an economic small-world network. Finally, a summary is given in Section \ref{sec:SUM}.

\section{Inhibitory Network of Suprathreshold FS Izhikevich Interneurons}
\label{sec:SWN}
A neural circuit in the major parts of the brain consists of a few types of excitatory principal cells and diverse types of inhibitory interneurons. By providing a synchronous oscillatory output to the principal cells, interneuronal networks play the role of the backbones of many brain rhythms \cite{Buz1,W_Review,Wang,Buz2}. We consider an inhibitory population of $N$ sparsely-coupled neurons equidistantly placed on a one-dimensional ring of radius $N/ 2 \pi$. As an element in our neural system, we choose the FS Izhikevich interneuron model \cite{Izhi1,Izhi2,Izhi3,Izhi4}. The population dynamics in this neural network is governed by the following set of ordinary differential equations:
\begin{eqnarray}
C\frac{dv_i}{dt} &=& k (v_i - v_r) (v_i - v_t) - u_i +I_{DC} +D \xi_{i} -I_{syn,i}, \label{eq:CIZA} \\
\frac{du_i}{dt} &=& a \{ U(v_i) - u_i \},  \;\;\; i=1, \cdots, N, \label{eq:CIZB}
\end{eqnarray}
with the auxiliary after-spike resetting:
\begin{equation}
{\rm if~} v_i \geq v_p,~ {\rm then~} v_i \leftarrow c~ {\rm and~} u_i \leftarrow u_i + d, \label{eq:RS}
\end{equation}
where
\begin{eqnarray}
U(v) &=& \left\{ \begin{array}{l} 0 {\rm ~for~} v<v_b \\ b(v - v_b)^3 {\rm ~for~} v \ge v_b \end{array} \right. , \label{eq:CIZC} \\
I_{syn,i} &=& \frac{J}{d_i^{in}} \sum_{j(\ne i)}^N w_{ij} s_j(t) (v_i - V_{syn}), \label{eq:CIZD}\\
s_j(t) &=& \sum_{f=1}^{F_j} E(t-t_f^{(j)}-\tau_l);~E(t) = \frac{1}{\tau_d - \tau_r} (e^{-t/\tau_d} - e^{-t/\tau_r}) \Theta(t). \label{eq:CIZE}
\end{eqnarray}
Here, the state of the $i$th neuron at a time $t$ is characterized by two state variables: the membrane potential $v_i$ and the recovery current $u_i$. In Eq.~(\ref{eq:CIZA}), $C$ is the membrane capacitance, $v_r$ is the resting membrane potential, and $v_t$ is the instantaneous threshold potential. After the potential reaches its apex (i.e., spike cutoff value) $v_p$, the membrane potential and the recovery variable are reset according to Eq.~(\ref{eq:RS}). The units of the capacitance $C$, the potential $v$, the current $u$ and the time $t$ are pF, mV, pA, and ms, respectively.

Unlike Hodgkin-Huxley-type conductance-based models, the Izhikevich model matches neuronal dynamics by tuning the parameters instead of matching neuronal electrophysiology. The parameters $k$ and $b$ are associated with the neuron's rheobase and input resistance, $a$ is the recovery time constant, $c$ is the after-spike reset value of $v$, and $d$ is the total amount of outward minus inward currents during the spike and affecting the after-spike behavior (i.e., after-spike jump value of $u$). Tuning these parameters, the Izhikevich neuron model may produce 20 of the most prominent neuro-computational features of cortical neurons \cite{Izhi1,Izhi2,Izhi3,Izhi4}. Here, we use the parameter values for the FS interneurons (which do not fire postinhibitory rebound spikes) in the layer 5 Rat visual cortex \cite{Izhi3}; $C=20,~v_r=-55,~v_t=-40,~v_p=25,~v_b=-55,~k=1,~a=0.2,~b=0.025,~c=-45,~d=0.$	

\begin{figure}
\includegraphics[width=0.8\columnwidth]{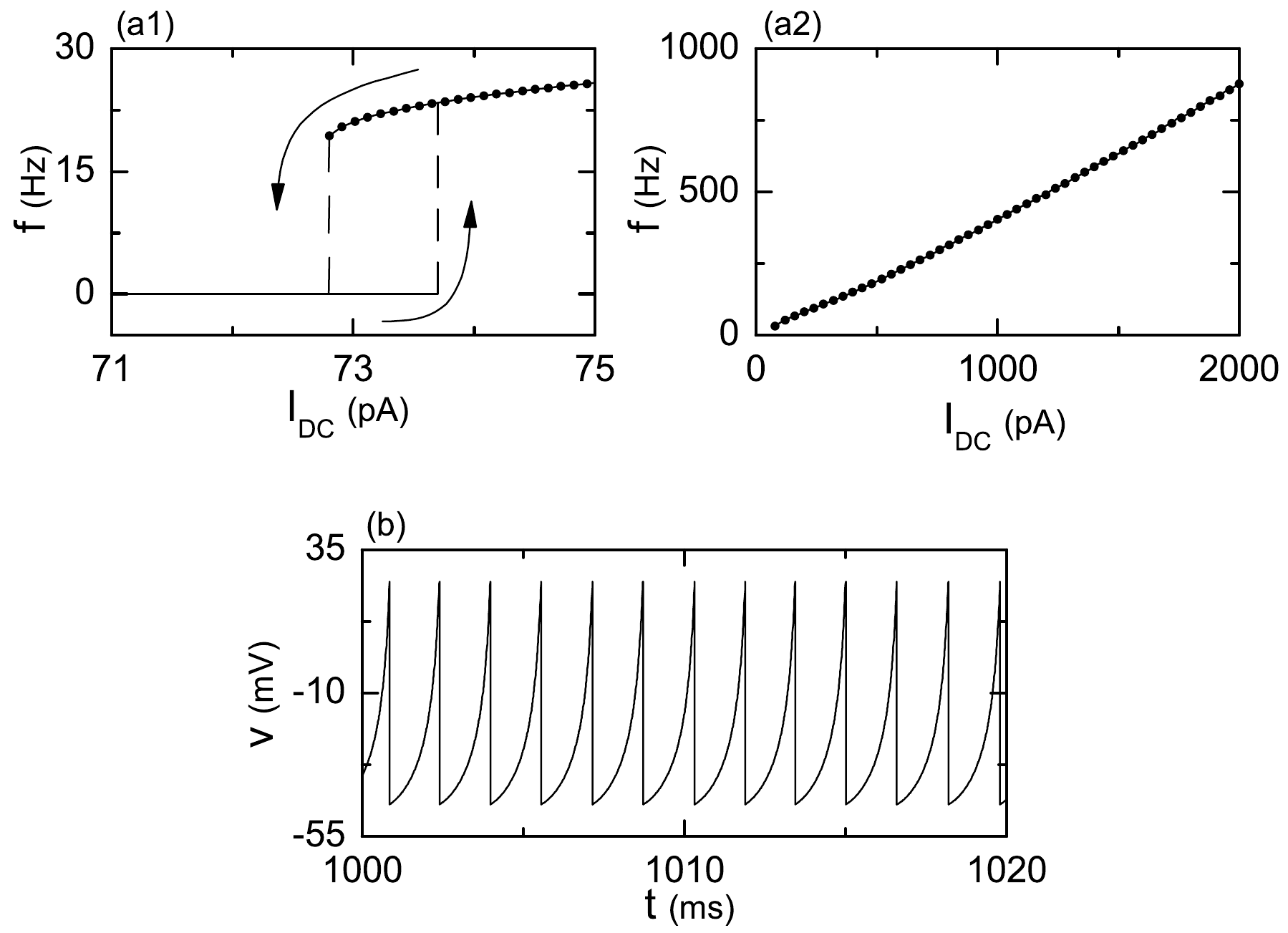}
\caption{Single FS Izhikevich interneuron for $D=0$. Plot of the mean firing rate $f$ versus the external DC current $I_{DC}$
(a1) near the transition point and (a2) in a large range of $I_{DC}$. (b) Time series of the membrane potential $v$
for $I_{DC}=1500$.
}
\label{fig:Single}
\end{figure}

Each Izhikevich interneuron is stimulated by using the common DC current $I_{DC}$ (measured in units of pA) and an independent Gaussian white noise $\xi_i$ [see the 3rd and the 4th terms in Eq.~(\ref{eq:CIZA})] satisfying $\langle \xi_i(t) \rangle =0$ and $\langle \xi_i(t)~\xi_j(t') \rangle = \delta_{ij}~\delta(t-t')$, where $\langle\cdots\rangle$ denotes the ensemble average. The noise $\xi$ is a parametric one that randomly perturbs the strength of the applied current $I_{DC}$, and its intensity is controlled by using the parameter $D$ (measured in units of ${\rm pA \cdot {ms}^{1/2}}$). In the absence of noise (i.e., $D=0$), the Izhikevich interneuron exhibits a jump from a resting state to a spiking state via subcritical Hopf bifurcation for $I_{DC,h}=73.7$ by absorbing an unstable limit cycle born via a fold limit cycle bifurcation for $I_{DC,l}=72.8$. Hence, the Izhikevich interneuron shows type-II excitability because it begins to fire with a non-zero frequency, as shown in Fig.~\ref{fig:Single}(a1)  \cite{Ex1,Ex2}. As $I_{DC}$ is increased from $I_{DC,h}$, the mean firing rate $f$ increases monotonically [see Fig.~\ref{fig:Single}(a2)]. Throughout this paper, we consider a suprathreshold case of $I_{DC}=1500$. For this case, the membrane potential $v$ oscillates very fast with $f=633$ Hz, as shown in Fig.~\ref{fig:Single}(b).

The last term in Eq.~(\ref{eq:CIZA}) represents the synaptic coupling of the network. $I_{syn,i}$ of Eq.~(\ref{eq:CIZD}) represents a synaptic current injected into the $i$th neuron. The synaptic connectivity is given by the connection weight matrix $W$ (=$\{ w_{ij} \}$) where  $w_{ij}=1$ if the neuron $j$ is presynaptic to the neuron $i$; otherwise, $w_{ij}=0$. Here, the synaptic connection is modeled by using both the conventional Erd\"{o}s-Renyi random graph and the Watts-Strogatz small-world network. Then, the in-degree of the $i$th neuron, $d_i^{in}$ (i.e., the number of synaptic inputs to the neuron $i$) is given by $d_i^{in} = \sum_{j(\ne i)}^N w_{ij}$. The fraction of open synaptic ion channels at time $t$ is denoted by $s(t)$. The time course of $s_j(t)$ of the $j$th neuron is given by a sum of delayed double-exponential functions $E(t-t_f^{(j)}-\tau_l)$ [see Eq.~(\ref{eq:CIZE})], where $\tau_l$ is the synaptic delay, and $t_f^{(j)}$ and $F_j$ are the $f$th spike and the total number of spikes of the $j$th neuron at time $t$, respectively. Here, $E(t)$ [which corresponds to contribution of a presynaptic spike occurring at time $0$ to $s(t)$ in the absence of synaptic delay] is controlled by the two synaptic time constants: synaptic rise time $\tau_r$ and decay time $\tau_d$, and $\Theta(t)$ is the Heaviside step function: $\Theta(t)=1$ for $t \geq 0$ and 0 for $t <0$. For the inhibitory GABAergic synapse (involving the $\rm{GABA_A}$ receptors), $\tau_l=1$ ms, $\tau_r=0.5$ ms, and $\tau_d=5$ ms \cite{Sparse6}. The coupling strength is controlled by the parameter $J$ (measured in units of $\rm \mu S$), and $V_{syn}$ is the synaptic reversal potential. Here, we use $V_{syn}=-80$ mV for the inhibitory synapse.

Numerical integration of Eqs.~(\ref{eq:CIZA})-(\ref{eq:CIZB}) is done using the Heun method \cite{SDE} (with the time step $\Delta t=0.01$ ms).
For each realization of the stochastic process, we choose a random initial point $[v_i(0),u_i(0),s_i(0)]$ for the $i$th $(i=1,\dots, N)$ neuron with uniform probability in the range of $v_i(0) \in (-50,-45)$, $u_i(0) \in (10,15)$, and $s_i(0) \in (0.0,0.02)$.

\section{Effect of Small-World Connectivity on Fast Sparsely Synchronized Rhythms}
\label{sec:SSR}
In this section, we study the effect of network architecture on fast sparsely synchronized rhythms with stochastic and intermittent neural discharges. For modeling the complex connectivity in neural systems, we first use the conventional Erd\"{o}s-Renyi random network of suprathreshold FS Izhikevich interneurons \cite{ER}, and study occurrence of population oscillatory states by varying the inhibition strength $J$ and the noise intensity $D$. But, in a real cortical circuit, synaptic connections are known to be neither regular nor completely random. Hence, we consider the Watts-Strogatz model for the small-world network which interpolates between the regular lattice and the random graph \cite{SWN1}, and investigate the effect of small-world connectivity on fast sparsely synchronized rhythms by varying the rewiring probability $p$ for fixed values of $J$ and $D$. Particularly, we search for an optimal fast sparse synchronization occurring at a minimal wiring cost in an economic small-world network.

\begin{figure}
\includegraphics[width=0.8\columnwidth]{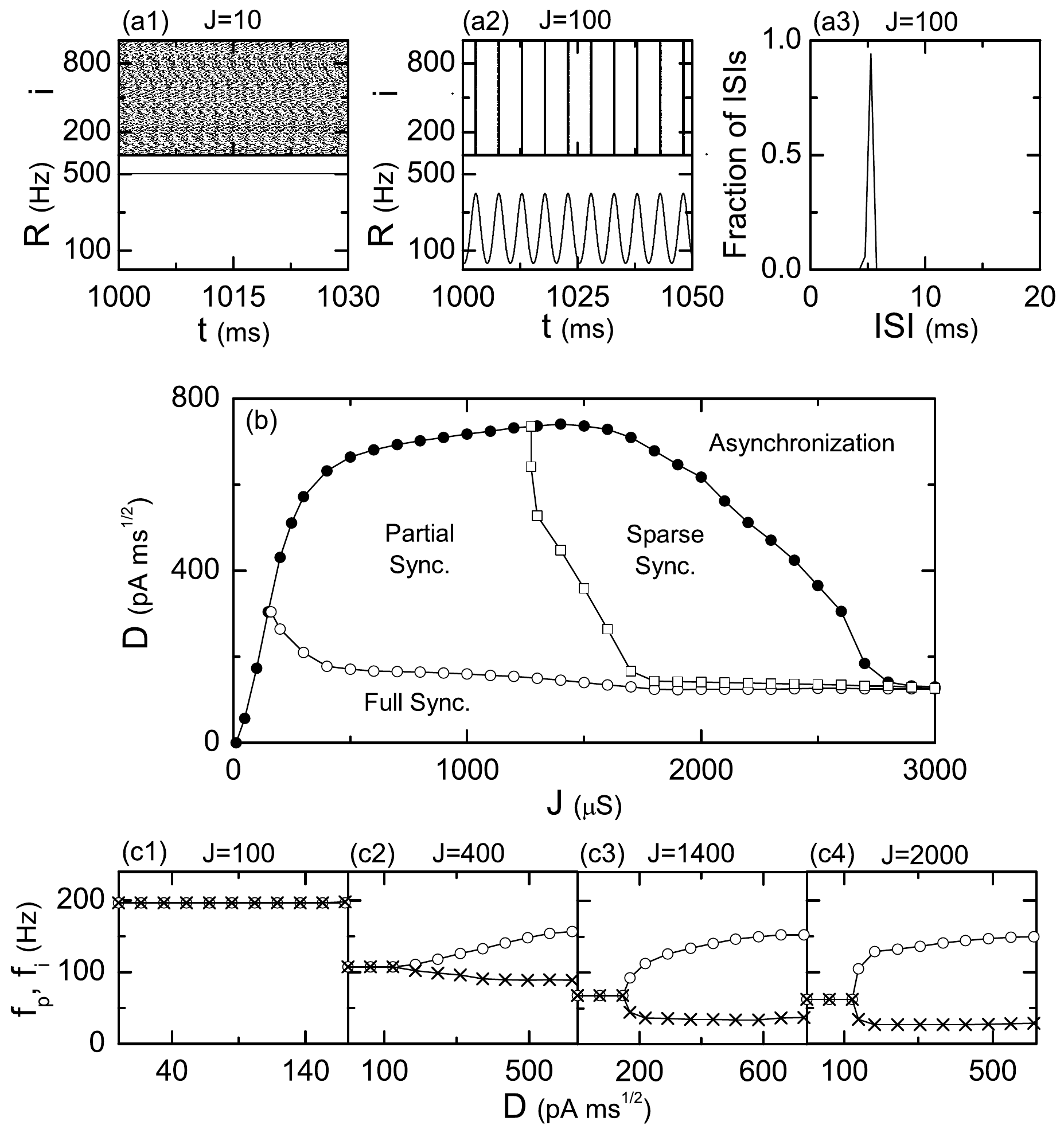}
\caption{Erd\"{o}s-Renyi random graph of $N$ $(=10^3)$ suprathreshold FS Izhikevich interneurons for $I_{DC}=1500$ and $M_{syn}=50$.
Raster plots of spikes and plots of the IPSR kernel estimate $R(t)$ versus $t$ for (a1) $J=10$ and (a2) $J=100$; the band width of the Gaussian kernel estimate is 1 ms. (a3) ISI histogram for $J=100$ (ISI histogram is composed of $5 \times 10^4$ ISIs and the bin size for the histogram is 0.5 ms). (b) State diagram in the $J-D$ plane. For the full synchronization, the individual frequency $f_i$ is the same as the population frequency $f_p$, while for the partial and sparse synchronization, $f_i$ is less than $f_p$. Particularly, the cases of $f_p > 4 \, f_i$ are referred to as the sparse synchronization. Plots of $f_p$ and $f_i$ versus $D$ for (c1) $J=100$, (c2) $J=400$, (c3) $J=1400$, and (c4) $J=2000$. Here, the circles and crosses denote $f_p$ and $f_i$, respectively.
}
\label{fig:SD}
\end{figure}

We first consider the conventional Erd\"{o}s-Renyi random graph of $N$ sparsely-connected suprathreshold FS Izhikevich interneurons equidistantly placed on a one-dimensional ring of radius $N/ 2 \pi$. A postsynaptic neuron $i$ receives a synaptic input from another presynaptic neuron $j$ with a connection probability $P_{syn}$ $(=M_{syn}/N)$, where $M_{syn}$ is the average number of synaptic inputs per neuron (i.e., $M_{syn} = \langle d_i \rangle$; $d_i$ is the number of synaptic inputs to the neuron $i$ and $\langle \cdots \rangle$ denotes an ensemble-average over all neurons). Here, we consider a sparse case of $M_{syn}=50$. By varying the inhibition intensity $J$ and the noise intensity $D$, we investigate occurrence of population synchronized states. In computational neuroscience, an ensemble-averaged global potential,
\begin{equation}
 V_G (t) = \frac {1} {N} \sum_{i=1}^{N} v_i(t),
\label{eq:GPOT}
\end{equation}
is often used for describing emergence of population neural synchronization.  However, to directly obtain $V_G$ in real experiments is very difficult. To overcome this difficulty, instead of $V_G$, we use an experimentally-obtainable IPSR which is often used as a collective quantity showing population behaviors \citep{W_Review,Sparse1,Sparse2,Sparse3,Sparse4,Sparse5,Sparse6}. The IPSR is obtained from the raster plot of neural spikes which is a collection of spike trains of individual neurons. Such raster plots of spikes, where population spike synchronization may be well visualized, are fundamental data in experimental neuroscience. For the synchronous case, ``stripes" (composed of spikes and indicating population synchronization) are found to be formed in the raster plot. Hence, for a synchronous case, an oscillating IPSR appears, while for an unsynchronized case the IPSR is nearly stationary. To obtain a smooth IPSR, we employ the kernel density estimation (kernel smoother) \cite{Kernel}. Each spike in the raster plot is convoluted (or blurred) with a kernel function $K_h(t)$ to obtain a smooth estimate of IPSR, $R(t)$:
\begin{equation}
R(t) = \frac{1}{N} \sum_{i=1}^{N} \sum_{s=1}^{n_i} K_h (t-t_{s}^{(i)}),
\label{eq:IPSRK}
\end{equation}
where $t_{s}^{(i)}$ is the $s$th spiking time of the $i$th neuron, $n_i$ is the total number of spikes for the $i$th neuron, and we use a Gaussian
kernel function of band width $h$:
\begin{equation}
K_h (t) = \frac{1}{\sqrt{2\pi}h} e^{-t^2 / 2h^2}, ~~~~ -\infty < t < \infty.
\label{eq:Gaussian}
\end{equation}
We first study the case of $D=0$. For small $J$, individual interneurons fire too fast to be synchronized. Figure \ref{fig:SD}(a1) shows the raster plot of spikes and the IPSR kernel estimate $R(t)$ for an unsynchronized case of $J=10$. Spikes in the raster plot are completely scattered  and hence $R(t)$ is  nearly stationary. However, as $J$ is increased mean firing rates, $f_i$, of individual interneurons decrease, and full synchronization occurs when $J$ passes a critical value $J^* (\simeq 12)$. For a synchronous case of $J=100$, clear stripes (composed of spikes and indicating population spike synchronization) are formed in the raster plot, and hence $R(t)$ shows regular oscillation with population frequency $f_p=197$ Hz, as shown in Fig.~\ref{fig:SD}(a2). The interspike interval (ISI) histogram with a single peak appearing at 5.1 ms is shown in Fig.~\ref{fig:SD}(a3), and hence individual neurons fire regularly with mean firing rate $f_i$ which is the same as $f_p$. Thus, complete full synchronization with $f_p=f_i$ occurs for $J=100$. We next consider the effect of noise on the full synchronization for a fixed $J$. As $D$ is increased, the full synchronization for $D=0$ evolves, depending on the values of $J$, and eventually desynchronization occurs when passing a critical value $D^*$. Figure \ref{fig:SD}(b) shows the state diagram in the $J-D$ plane. For the full synchronization, mean firing rates, $f_i$, of individual neurons are the same as the population frequency $f_p$, while for the partial and sparse synchronization, $f_i$ is less than $f_p$ (i.e., individual neurons fire at lower rates than the population frequency). For the sparsely synchronized cortical rhythms, $f_p:f_i \sim 4:1$ \cite{Sparse1,Sparse2,Sparse3,Sparse4}. Hence, when the population frequency is much higher than the mean firing rate of individual interneurons ($f_p > 4\, f_i$), the synchronization will be referred to as sparse synchronization. Plots of $f_p$ and $f_i$ versus $D$ are shown in Figs.~\ref{fig:SD}(c1)-\ref{fig:SD}(c4) for $J=$ 100, 400, 1400, and 2000. For small $J$ ($J^* < J < 162)$, only the full synchronization occurs because $f_p = f_i$ (e.g., see the case of $J=100$). However, for $J>162$, the full synchronization is developed into partial synchronization at some threshold value $D_{th}$ via pitchfork-like bifurcations (e.g., see the cases of $J=400$ 1400, and 2000). With increasing $J$, the difference between $f_p$ and $f_i$ increases abruptly when passing $D_{th}$. For $J>1275$, the partial synchronization evolves into sparse synchronization with $f_p > 4\,f_i$ (e.g., see the cases of $J=1400$ and 2000).

\begin{figure}
\includegraphics[width=0.8\columnwidth]{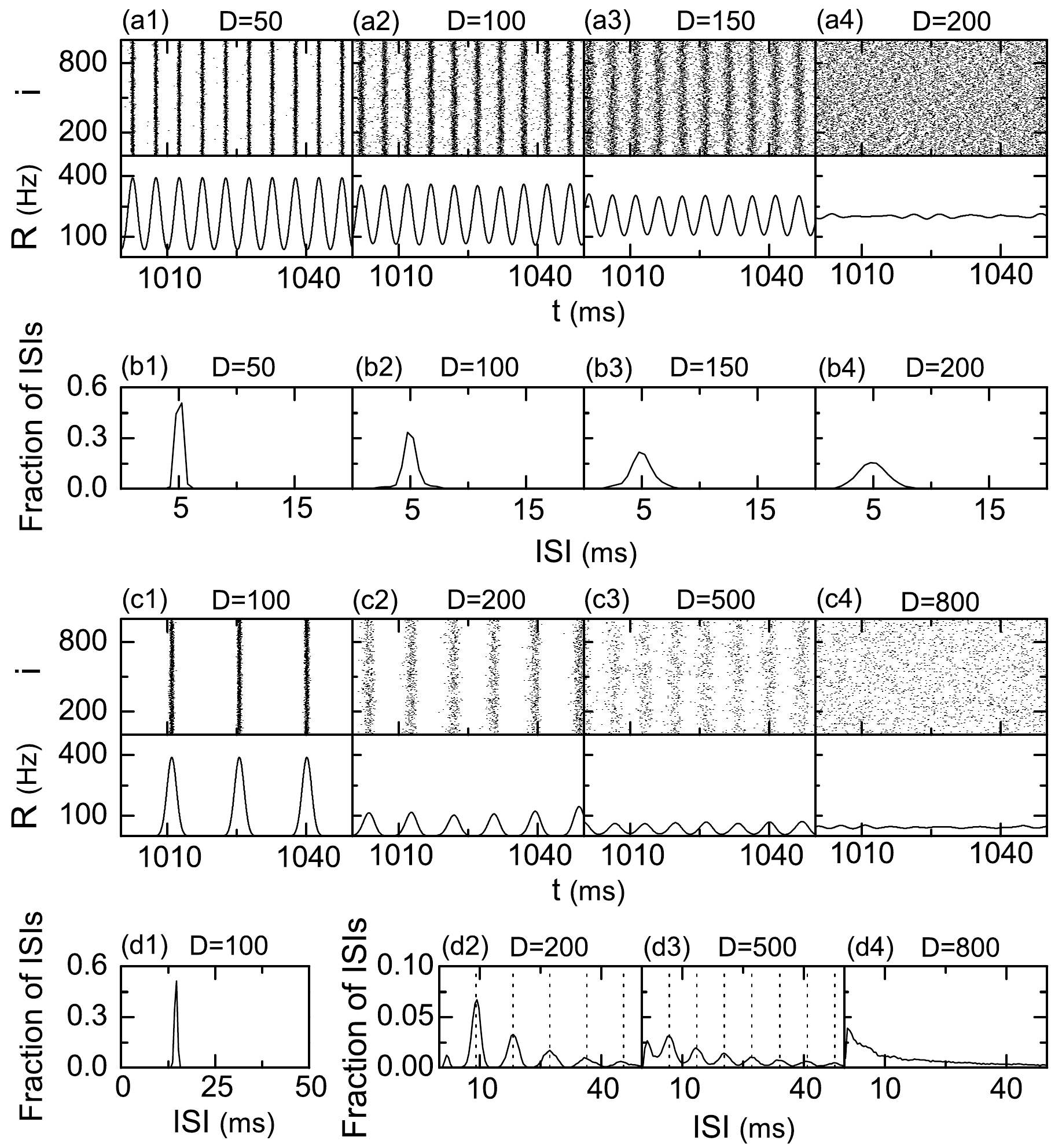}
\caption{Erd\"{o}s-Renyi random graph of $N$ $(=10^3)$ suprathreshold FS Izhikevich interneurons for $I_{DC}=1500$ and $M_{syn}=50$.
For $J=100$, raster plots of spikes and plots of the IPSR kernel estimate $R(t)$ versus $t$ in (a1)-(a4), and ISI histograms in (b1)-(b4) for various values of $D$. For $J=1400$, raster plots of spikes and plots of $R(t)$ versus $t$ in (c1)-(c4), and ISI histograms in (d1)-(d4) for various values of $D$; vertical dotted lines denote integer multiples of the global period $T_G$ [$\simeq 9.1$ ms in (d2) and 6.8 ms in (d3)] of $R(t)$. The band width of the Gaussian kernel estimate is 1 ms. Each ISI histogram is composed of $5 \times 10^4$ ISIs and the bin size for the histogram is 0.5 ms.
}
\label{fig:RA}
\end{figure}

For further understanding of the full and the partial synchronization, we present two explicit examples showing how the full synchronization is evolved into unsynchronized states as $D$ is increased. For $J^*<J<162$, the full synchronization for $D=0$ develops directly into an unsynchronized state without any other type of intermediate synchronization stage. As an example consider the case of $J=100$. The raster plots and the IPSR kernel estimate $R(t)$ for various values of $D$ are given in Figs.~\ref{fig:RA}(a1)-\ref{fig:RA}(a4). As $D$ is increased, stripes of spikes in the raster plot become more and more smeared, and hence the amplitudes of $R(t)$ become smaller (i.e., the pacing degree of spikes decreases). When passing a critical value $D^* \simeq 173$, stripes become overlapped and $R(t)$ becomes nearly stationary. Thus, a transition to an unsynchronized state occurs [e.g., see Fig.~\ref{fig:RA}(a4)]. The ISI distributions for various values of $D$ are also shown in Figs.~\ref{fig:RA}(b1)-\ref{fig:RA}(b4). As $D$ is increased from 0, the height of the ISI histogram becomes smaller and its width becomes wider. During this process, $f_p=f_i \simeq 197$ Hz, as shown in Fig.~\ref{fig:SD}(c1). Thus, as $D$ passes a critical value $D^*$, a direct transition from full synchronization to an unsynchronized state occurs for $J=100$. For $J> 162$, with increasing $D$ the full synchronization for $D=0$ evolves into the partial synchronization with $f_p > f_i$. As an example, we consider the case of $J=1400$. Figures \ref{fig:RA}(c1)-\ref{fig:RA}(c4) show the raster plots and the IPSR kernel estimate $R(t)$ for various values of $D$. For $0<D<D_{th} (\simeq 144)$, full synchronization with $f_p = f_i$ occurs [see Fig.~\ref{fig:SD}(c3)]. As $D$ is increased from 0 to $D_{th}$, the degree of full synchronization decreases because the stripes of the raster plot become smeared, and hence the amplitude of $R(t)$ also becomes smaller [see Fig.~\ref{fig:RA}(c1)]. As in the case of $J=100$, the width of the ISI histogram becomes wider due to decrease in the pacing degree [see Fig.~\ref{fig:RA}(d1)]. However, for $D>144$, partial synchronization with $f_p > f_i$ appears via a pitchfork-like bifurcation, as shown in Fig.~\ref{fig:SD}(c3). As $D$ is increased from $D=144$, the interval between stripes of the raster plot becomes smaller, and hence the population frequency $f_p$ increases [see Fig.~\ref{fig:RA}(c2)]. For this case, the stripes of the raster plot become smeared, and hence the pacing degree of spikes decreases. Furthermore, the density of stripes becomes smaller because smaller fraction of total neurons fire in each stripes. Thus, with increasing $D$ from $D=144$ both the pacing and the occupation degrees of spikes decrease, and consequently a large decrease in the amplitude of $R(t)$ occurs. In contrast to the case of full synchronization, the ISI histogram has multiple peaks appearing at multiples of the period $T_G$ of $R(t)$ [see Figs.~\ref{fig:RA}(d2)]. Similar skipping phenomena of spikings (characterized with multi-peaked ISI histograms) have also been found in networks of coupled inhibitory neurons in the presence of noise where noise-induced hopping from one cluster to another one occurs \cite{GR}, in single noisy neuron models exhibiting stochastic resonance due to a weak periodic external force \cite{Longtin1,Longtin2}, and in inhibitory networks of coupled subthreshold neurons showing stochastic spiking coherence \cite{Kim1,Kim2,Kim3}. Stochastic spike skipping in coupled systems is a collective effect because it occurs due to a driving by a coherent ensemble-averaged synaptic current, in contrast to the single case driven by a weak periodic force where stochastic resonance occurs. Due to this stochastic spike skipping, partial occupation occurs in the stripes of the raster plot. Thus, the mean firing rates $f_i$ of individual interneurons become less than the population frequency $f_p$, and hence partial synchronization occurs. Particularly, for $D>448$, sparse synchronization with $f_p > 4\, f_i$ appears. As $D$ is further increased from $D=448$, both the pacing and the occupation degrees of spikes (seen in the raster plot) decrease and multiple peaks in the ISI histogram overlap [see Figs.~\ref{fig:RA}(c3) and \ref{fig:RA}(d3)]. Eventually, when passing a critical value $D^* \simeq 741$, an unsynchronized state appears (e.g. see the case of $D=800)$.

As shown in the state diagram of Fig.~\ref{fig:SD}(b), fast sparsely synchronized rhythms appear when both the inhibition strength and the noise intensity are strong in the Erd\"{o}s-Renyi random graph of suprathreshold FS Izhikevich interneurons. For random connectivity, the average path length is short due to long-range connections, and hence global efficiency of information transfer becomes high \cite{Eff1,Eff2}. However, unlike the regular lattice, the random network has poor clustering, and it becomes non-economic due to appearance of short-range and long-range connections with equal probability \cite{Sporns,Buz2}. Real synaptic connectivity is known to have complex topology which is neither regular nor completely random \cite{Sporns,Buz2,CN1,CN2,CN3,CN4,CN5,CN6}. To study the effect of network structure on fast sparsely synchronized oscillations, we consider the Watts-Strogatz model for small-world networks which interpolates between regular lattice and random graph via rewiring \cite{SWN1}. By varying the rewiring probability $p$ from local to long-range connection, we investigate the effect of small-world connectivity on fast sparse synchronization for fixed values of $J=1400$ and $D=500$. We start with a directed regular ring lattice with $N$ suprathreshold FS Izhikevich interneurons where each Izhikevich interneuron is coupled to its first $M_{syn}$ neighbors ($M_{syn}/2$ on either side) via outward synapses, and rewire each outward connection at random with probability $p$ such that self-connections and duplicate connections are excluded. As in the above random case, we consider a sparse but connected network with a fixed value of $M_{syn}= 50$. Then, we can tune the network between regularity $(p=0)$ and randomness $(p=1)$; the case of $p=1$ corresponds to the above Erd\"{o}s-Renyi random graph. In this way, we investigate emergence of fast sparsely synchronized rhythm in the directed Watts-Strogatz small-world network of $N$ suprathreshold FS Izhikevich interneurons by varying the rewiring probability $p$ for $J=1400$ and $D=500$.

The topological properties of the small-world connectivity has been well characterized in terms of the clustering coefficient and the average path length \cite{SWN1}. The clustering coefficient, denoting the cliquishness of a typical neighborhood in the network, characterizes the local efficiency of information transfer, while the average path length, representing the typical separation between two vertices in the network, characterizes the global efficiency of information transfer. The regular lattice for $p=0$ is highly clustered but large world where the average path length grows linearly with $N$, while the random graph for $p=1$ is poorly clustered but small world where the average path length grows logarithmically with $N$ \cite{SWN1}. As soon as $p$ increases from zero, the average path length decreases dramatically, which leads to occurrence of a small-world phenomenon which is popularized by the phrase of the ``six degrees of separation'' \cite{SDS1,SDS2}. However, during this dramatic drop in the average path length, the clustering coefficient remains almost constant at its value for the regular lattice. Consequently, for small $p$ small-world networks with short path length and high clustering emerge \cite{SWN1}.

\begin{figure}
\includegraphics[width=\columnwidth]{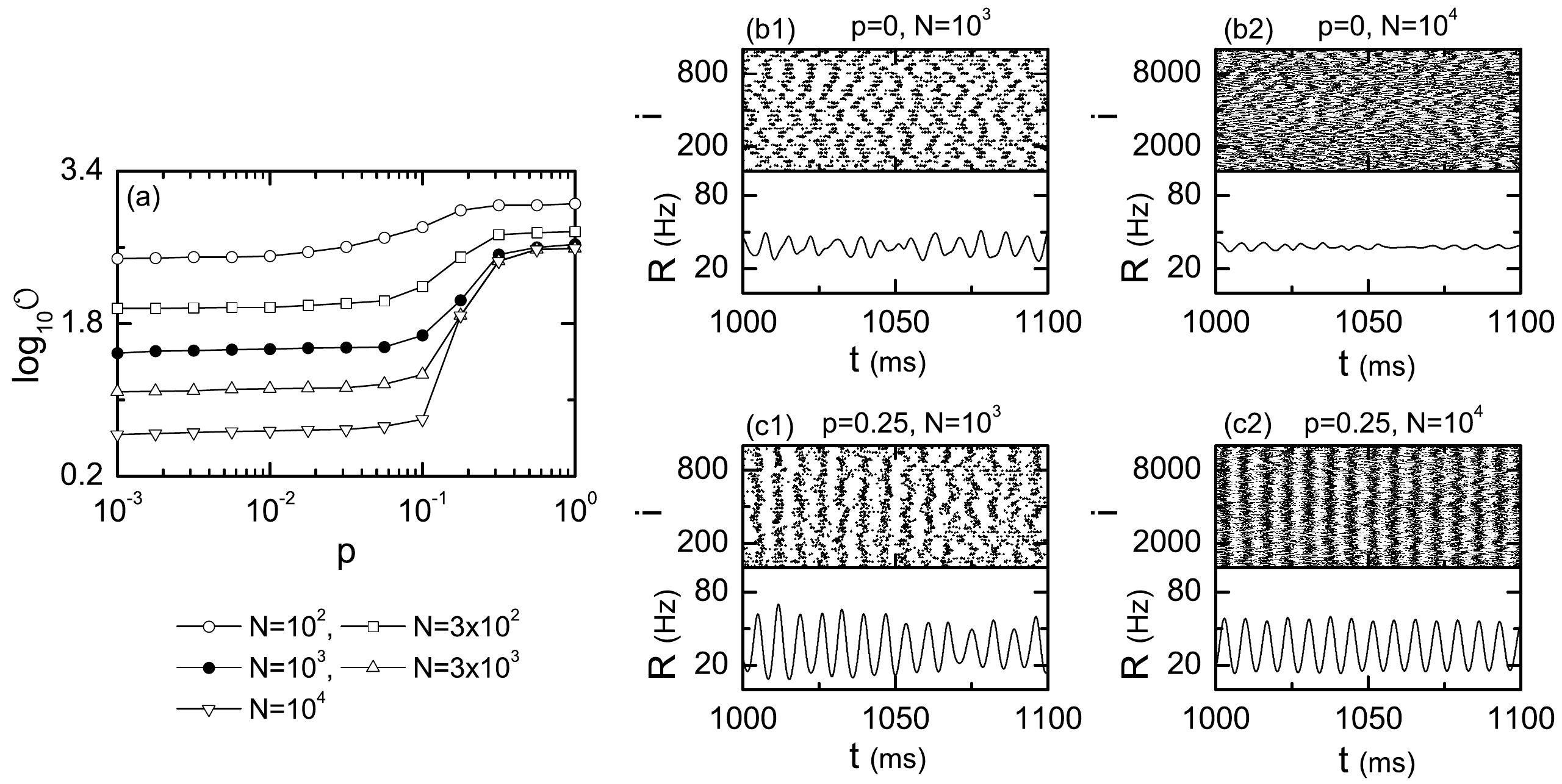}
\caption{Watts-Strogatz small-world network of suprathreshold FS Izhikevich interneurons for $I_{DC}=1500$, $M_{syn}=50$, $J=1400$, and $D=500$.
(a) Plots of $\log_{10}\cal{O}$ versus $p$. Unsynchronized state for $p=0$: raster plots of spikes and plots of the IPSR kernel estimate $R(t)$ versus $t$ for (b1) $N=10^3$ and (b2) $N=10^4$. Synchronized state for $p=0.25$: raster plots of spikes and plots of the IPSR kernel estimate $R(t)$ versus $t$ for (c1) $N=10^3$ and (c2) $N=10^4$. The band width of the Gaussian kernel estimate is 1 ms.
}
\label{fig:Order}
\end{figure}

As is well known, a conventional order parameter, based on the ensemble-averaged global potential $V_G$, is often used for describing transition from asynchrony to synchrony in computational neuroscience \cite{Order1,Order2,Order3}. Recently, instead of $V_G$, we used an experimentally-obtainable IPSR kernel estimate $R(t)$, and developed a realistic order parameter, which may be applicable in both the computational and the experimental neuroscience \cite{RM}. The mean square deviation of $R(t)$,
\begin{equation}
{\cal{O}} \equiv \overline{(R(t) - \overline{R(t)})^2},
 \label{eq:Order}
\end{equation}
plays the role of an order parameter $\cal{O}$. (Here the overbar represents the time average.) The order parameter may be regarded as a thermodynamic measure because it concerns just the macroscopic IPSR kernel estimate $R(t)$ without any consideration between $R(t)$ and microscopic individual spikes. In the thermodynamic limit of $N \rightarrow \infty$, the order parameter $\cal{O}$ approaches a non-zero (zero) limit value for the synchronized (unsynchronized) state. Figure \ref{fig:Order}(a) shows a plot of the order parameter versus the rewiring parameter $p$. For $p < p_c^*$ $(\simeq 0.12$), unsynchronized states exist because the order parameter $\cal{O}$ tends to zero as $N \rightarrow \infty$. As $p$ passes the critical value $p_c^*$, a transition to synchronization occurs because the values of $\cal {O}$ become saturated to non-zero limit values for $N \geq 3 \cdot 10^3$. These synchronized states seem to appear because global efficiency of information transfer between distant neurons for $p>p_c^*$ becomes enough for occurrence of population synchronization. Here we present two explicit examples for the synchronized and unsynchronized states. First, we consider the population state in the regular lattice for $p=0$. As shown in Fig.~\ref{fig:Order}(b1) for $N=10^3$, the raster plot shows a zigzag pattern intermingled with inclined partial stripes of spikes with diverse inclinations and widths, and $R(t)$ is composed of coherent parts with regular large-amplitude oscillations and incoherent parts with irregular small-amplitude fluctuations. For $p=0$, the clustering coefficient is high, and hence partial stripes (indicating local clustering of spikes) seem to appear in the raster plot of spikes. As $N$ is increased to $10^4$, partial stripes become more inclined from the vertical, and hence spikes become more difficult to keep pace with each other. As a result, $R(t)$ shows noisy fluctuations with smaller amplitudes, as shown in Fig.~\ref{fig:Order}(b2). Hence the population state for $p=0$ seems to be unsynchronized because $R(t)$ tends to be nearly stationary as $N$ increases to the infinity. As $p$ is increased from 0, long-range short-cuts begin to appear, and hence average path length becomes shorter. Eventually, when passing the critical value $p_c^*$, synchronized population state emerges because of sufficient global efficiency of information transfer between distant neurons, which will be discussed below in more details. As a second example, we consider a synchronized case of $p=0.25$. For $N=10^3$, the degree of zigzagness for partial stripes in the raster plot is much reduced when compared with the $p=0$ case, and hence $R(t)$ shows a regular oscillation, as shown in Fig.~\ref{fig:Order}(c1). Its amplitudes are much larger than that for the case of $p=0$, although there is a little variation in the amplitude. As $N$ is increased to $N=10^4$, $R(t)$ shows regular oscillations, and the amplitudes in each oscillating cycle are nearly the same, in contrast to the case of $N=10^3$ [see Fig.~\ref{fig:Order}(c2)]. Hence, $R(t)$ displays more regular oscillations with nearly the same amplitudes for $N=10^4$. Consequently, the population state for $p=0.25$ seems to be synchronized because $R(t)$ tends to show regular oscillations as $N$ goes to the infinity.

\begin{figure}
\includegraphics[width=0.8\columnwidth]{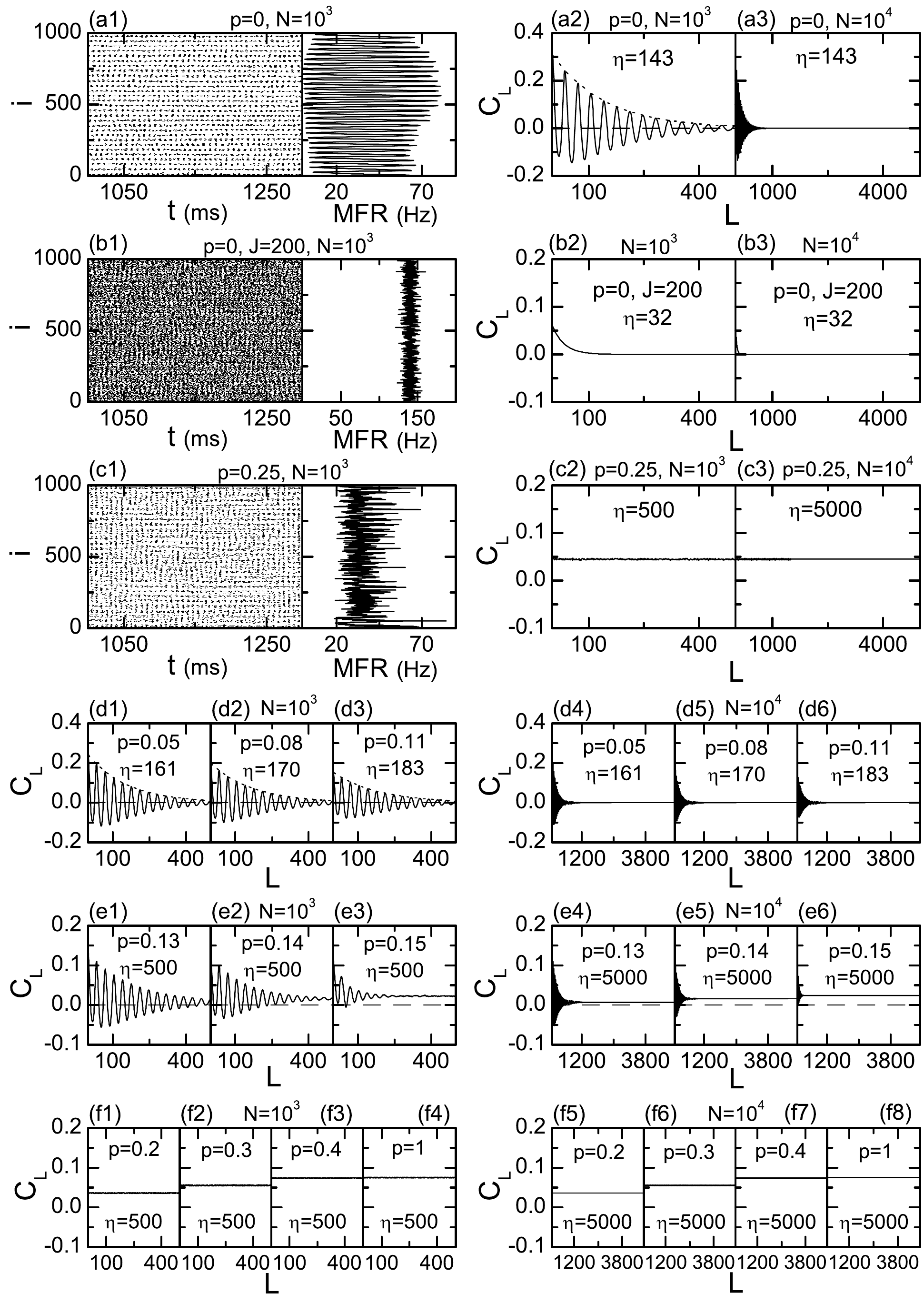}
\linespread{1.2}
\caption{Watts-Strogatz small-world network of suprathreshold FS Izhikevich interneurons for $I_{DC}=1500$, $M_{syn}=50$, and $D=500$. Unsynchronized state for $p=0$ and $J=1400$: (a1) raster plots of spikes and MFR distribution of individual neurons for $N=10^3$ and the spatial correlation function $C_L$ for $N=$ (a2) $10^3$ and (a3) $10^4$. For this case of $J=1400$, the MFR distribution exhibits spatially modulated oscillation. Unsynchronized state for $p=0$ and $J=200$: (b1) raster plots of spikes and MFR distribution of individual neurons for $N=10^3$ and the spatial correlation function $C_L$ for $N=$ (b2) $10^3$ and (b3) $10^4$. The MFR distribution for $J=200$ shows noisy fluctuation around the average MFR ($\simeq 140$ Hz) without any regular spatial oscillation. Hereafter, the values of $J$ in (c1)-(f8) are fixed at $J=1400$. Synchronized state for $p=0.25$: (c1) raster plots of spikes and MFR distribution of individual neurons for $N=10^3$ and the spatial correlation function $C_L$ for $N=$ (c2) $10^3$ and (c3) $10^4$. Spatial correlation functions $C_L$ for unsynchronized states for various values of $p$ when (d1)-(d3) $N=10^3$ and (d4)-(d6) $10^4$. Spatial correlation functions $C_L$ for synchronized states for various values of $p$ when (e1)-(e3) $N=10^3$ and (e4)-(e6) $10^4$. Flat spatial correlation functions $C_L$ for synchronized states for various values of $p$ when (f1)-(f4) $N=10^3$ and (f5)-(f8) $10^4$. The band width of the Gaussian kernel estimate is 1 ms. The temporal cross-correlation $C_{i,j}(\tau)$ is obtained through average over 20 realizations, and the number of data used for the calculation of each temporal cross-correlation function $C_{i,j}(\tau)$ is $2\times10^{4}$.}
\label{fig:Corr}
\end{figure}

In order to further understand the above transition to global spike synchronization, we investigate the effect of geometrical long-range connections on the dynamical cross-correlations between neuronal pairs. As examples, we reconsider the same cases of $p=0$ and 0.25 as in Fig.~\ref{fig:Order}.
Figure \ref{fig:Corr}(a1) shows a raster plot of spikes in the longer time interval [where the horizontal structure may be more clearly seen than the vertical zigzag structure when compared with the raster plot (in the shorter time interval) in Fig.~\ref{fig:Order}(b1)] and the distribution of the mean firing rates (MFRs) of individual neurons for the case of $p=0$ (without long-range connections). Dense and sparse horizontal stripes of spikes appear alternatively in the raster plot with the average spatial period $\tau_L$ $(\simeq 36)$. In accordance to this structure of the horizontal stripes, the MFR distribution also shows a regular oscillation with large amplitude around the average MFR ($\simeq 33$ Hz): the peak (valley) regions of the MFR correspond to the centers of the dense (sparse) stripes. This spatial oscillation of the MFR distribution affects the behavior of the spatial cross-correlations between neuronal pairs. For obtaining dynamical pair cross-correlations, each spike train of the $i$th neuron is convoluted with a Gaussian kernel function $K_h(t)$ of band width $h$ to get a smooth estimate of instantaneous individual spike rate (IISR), $r_i(t)$:
\begin{equation}
r_i(t) = \sum_{s=1}^{n_i} K_h (t-t_{s}^{(i)}),
\label{eq:IISR}
\end{equation}
where $t_{s}^{(i)}$ is the $s$th spiking time of the $i$th neuron, $n_i$ is the total number of spikes for the $i$th neuron, and $K_h(t)$ is given in Eq.~(\ref{eq:Gaussian}). Then, the normalized temporal cross-correlation function $C_{i,j}(\tau)$ between the IISRs $r_i(t)$ and $r_j(t)$ of the $(i,j)$ neuronal pair is given by:
\begin{equation}
C_{i,j}(\tau) = \frac{\overline{\Delta r_i(t+\tau) \Delta r_j(t)}}{\sqrt{\overline{\Delta r^2_i(t)}}\sqrt{\overline{\Delta r^2_j(t)}}},
\end{equation}
where $\Delta r_i(t) = r_i(t) - \overline{r_i(t)}$ and the overline denotes the time average.
We now introduce the spatial cross-correlation $C_L$ ($L=1,...,N/2)$ between neuronal pairs separated by a spatial distance $L$ through average of all the temporal cross-correlations between $r_i(t)$ and $r_{i+L}(t)$ $(i=1,...,N)$ at the zero-time lag:
\begin{equation}
C_L = \frac{1}{N} \sum_{i=1}^{N} C_{i, i+L}(0) ~~~~ {\rm for~} L=1, \cdots, N/2.
\end{equation}
Figure \ref{fig:Corr}(a2) shows the plot of the spatial cross-correlation $C_L$ versus $L$ for $N=10^3$ in the case of $p=0$. We note that the spatial cross-correlation function $C_L$ exhibits a ``damped'' oscillation with respect to $L$. Local maxima (minima) of $C_L$ come from cross-correlations when the distances $L$ between neuronal pairs are multiples (odd multiples) of the spatial period $\tau_L$ (half-period $\tau_L/2$) of the MFR. In this way, $C_L$
makes an oscillatory decay to zero. To obtain the exponential decay rate of $C_L$, the maximal envelope (denoted by a dotted line) may be well fitted with an exponential function with a characteristic correlation $\eta$:
\begin{equation}
C_L = A \cdot e^{-L/\eta}; ~~~~ A=0.31, \eta=143.
\end{equation}
Then, one can think that the whole system is composed of independent partially synchronized blocks of size $\eta$. To examine occurrence of population synchronization, we increase the number of neurons as $N=10^4$. For this case, the spatial cross-correlation function $C_L$ in Fig.~\ref{fig:Corr}(a3) also show a damped oscillation, and it decays to zero. Then, the maximal envelope also is well fitted with the same exponential function ($C_L = A \cdot e^{-L/\eta}; A=0.31, \eta=143 $). Hence, the correlation length $\eta$ remains unchanged, although the system size is increased 10 times. As a criterion for occurrence of synchronization, we introduce a normalized correlation length $\tilde{\eta}$:
\begin{equation}
     \tilde{\eta} = \frac {\eta} {N},
\label{eq:NCC}
\end{equation}
which represents the ratio of the correlation length $\eta$ to the system size $N$. As $N$ is increased, $\tilde{\eta}$ tends to zero [i.e., the relative size of partially synchronized blocks (when compared to the whole system size) tends to zero]. Consequently, no global synchronization occurs for the case of $p=0$.
Although the population dynamical state for $p=0$ is unsynchronized, it seems to be worth noting the spatially modulated MFR distribution which exhibits regular spatial oscillation for $J=1400$. To examine appearance of spatially modulated phase, we decrease the inhibitory coupling strength $J$, and find spatially non-modulated phase for small $J$ less than a critical value $J_c$ $(\sim 540)$. Figure \ref{fig:Corr}(b1) shows an example of spatially non-modulated phase for $J=200$ where the MFR distribution exhibits a noisy fluctuation around the average MFR ($\simeq 140$ Hz) without any regular oscillation. The average MFR for $J=200$ is much larger than that for $J=1400$ because the inhibition strength $J$ is decreased. In this case of non-modulated phase, the spatial cross-correlation function $C_L$ exhibits a direct exponential decay to zero without any oscillation, as shown in Figs.~\ref{fig:Corr}(b2) and \ref{fig:Corr}(b3), unlike the case of modulated phase for $J=1400$ where $C_L$ shows an oscillatory decay to zero. Furthermore, the correlation length for $J=200$ is $\eta \simeq 32$, independently of the system size $N$, which is smaller than that for the case of $J=1400$ because the decay rate to zero for $J=200$ is faster than that for $J=1400$. Although the topological clustering coefficient is high for $p=0$, no distinct dynamical clustering seems to occur when the inhibitory coupling strength $J$ is small. The clustering effect seems to be distinctly evident only when $J$ becomes sufficiently strong. Hence, when $J$ passes the critical value $J_c$, this type of  spatially modulated phase seems to emerge spontaneously. Further in-depth study of this transition to spatially modulated phase is beyond our present subject, and it will be done in a future work. Then, we return to the original subject on the effect of small-world connectivity for a fixed value of $J=1400$, and consider another case of $p=0.25$ where long-range connections appear. The raster plot of spikes and the MFR distribution are shown in Fig.~\ref{fig:Corr}(c1). Some of dense horizontal stripes become smeared and overlap vertically with neighboring dense stripes, and then some parts of sparse horizontal stripes between merging dense stripes disappear. The MFR distribution shows irregular oscillation with relatively small dispersion about the average MFR ($\simeq 33$ Hz)(when compared with the MFR distribution for $p=0$). Unlike the case of $p=0$, the spatial correlation function $C_L$ for $N=10^3$ becomes nearly non-zero constant $(\simeq 0.045)$ in the whole range of $L$, as shown in Fig.~\ref{fig:Corr}(C2), and hence the correlation length $\eta$ becomes $N/2$ (=500) covering the whole system (note that the maximal distance between neurons is $N/2$ because of the ring architecture on which neurons exist). Consequently, the whole system is composed of just one single synchronized block (i.e., global synchronization occurs in the whole system), in contrast to the case of $p=0$.  For $N=10^4$, the flatness of $C_L$ in Fig.~\ref{fig:Corr}(C3) also extends to the whole range ($L=N/2=5000$), and the correlation length becomes $\eta=5000$, which also covers the whole system. Hence, the number of long-range connections for $p=0.25$ seems to become enough to make the correlation length $\eta$ cover the whole system, independently of $N$. Consequently, as $N$ is increased, the normalized correlation length $\tilde{\eta}$ has a non-zero limit value, $1/2$, and global synchronization emerges in the whole population, in contrast to the case of $p=0$. In addition to the above cases of $p=0$ and $0.25$, we also make an extensive dynamical-correlation analysis for several values of $p$ in the subcritical and the supercritical subregions of $p$. Figures \ref{fig:Corr}(d1)-\ref{fig:Corr}(d6) show the spatial correlation functions $C_L$ in the subcritical cases of $p=0.05$, 0.08, and 0.11 [less than the critical values $p^*_c$ $(\simeq 0.12$)]. For $N=10^3$, $C_L$  makes an oscillatory decay to zero. As $p$ is increased, the decay rate becomes slower, and hence the correlation length $\eta$ becomes longer, as shown in Figs.~\ref{fig:Corr}(d1)-\ref{fig:Corr}(d3). When the system size is increased to $N=10^4$, $C_L$ also exhibits an oscillatory decay to zero in the same way as the case of $N=10^3$ [see Figs.~\ref{fig:Corr}(d4)-\ref{fig:Corr}(d6)]. Hence, the correlation length $\eta$ becomes the same, independently of $N$, and  the normalized correlation length $\tilde{\eta}$ $(=\eta/N)$ tends to zero in the thermodynamic limit of $N \rightarrow \infty$ due to insufficient number of long-range connections. As a result, no global synchronization occurs for the subcritical case. On the other hand, in the supercritical case of $p=0.13$, 0.14, and 0.15 (larger than $p^*_c$), $C_L$ seems to make an oscillatory decay to a non-zero limit, as shown in Figs.~\ref{fig:Corr}(e1)-\ref{fig:Corr}(e3), and the convergence rate to non-zero limit becomes faster with increasing $p$. As $N$ is increased, this kind of convergence to non-zero limit in $C_L$ may be more clearly seen [e.g., see Figs.~\ref{fig:Corr}(e4)-\ref{fig:Corr}(e6) for $N=10^4$]. As $p$ is further increased, the oscillatory part in $C_L$ disappears gradually, and then $C_L$ becomes nearly positive constant, as shown in Figs.~\ref{fig:Corr}(f1)-\ref{fig:Corr}(f8) for both $N=10^3$ and $10^4$. Hence, unlike the subcritical case, with increasing $N$ the correlation length $\eta$ increases as $N/2$, covering the whole system, and the normalized correlation length $\tilde{\eta}$ has a non-zero limit value, $1/2$, in the thermodynamic limit of $N \rightarrow \infty$. Consequently, global population synchronization occurs for the supercritical case because $\eta$ covers the whole system thanks to sufficient number of long-range connections. We also note that, with increasing $p$ the constant values of $C_L$ increase, as shown in Figs.~\ref{fig:Corr}(f1)-\ref{fig:Corr}(f8). However, their values become saturated for $p = p_{max} (\sim 0.4)$ (i.e., $C_L = 0.075$ for $p \geq p_{max}$), because long-range connections which appear up to $p_{max}$ seem to play sufficient role for obtaining the maximal degree of pair cross-correlations.

\begin{figure}
\includegraphics[width=\columnwidth]{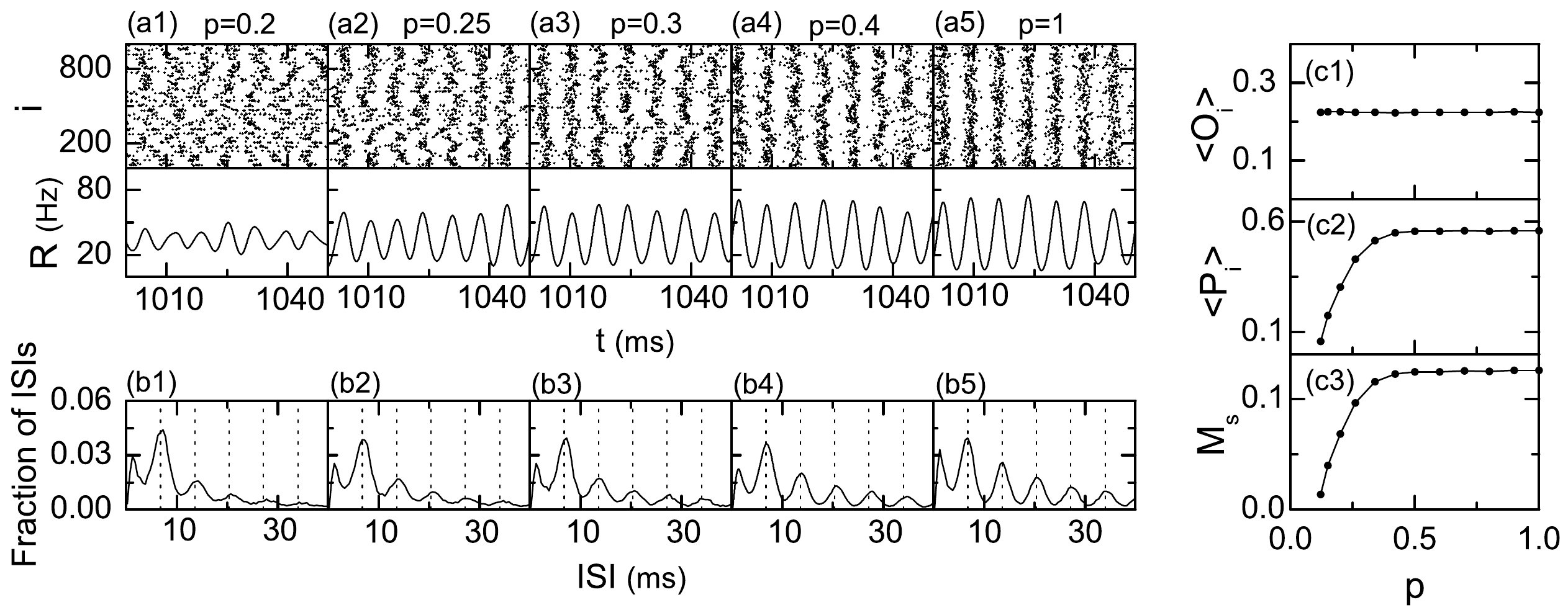}
\caption{Watts-Strogatz small-world network of suprathreshold FS Izhikevich interneurons for $N=10^3$, $I_{DC}=1500$, $M_{syn}=50$, $J=1400$, and $D=500$. Fast sparsely synchronized states for various values of $p$: raster plots of neural spikes and plots of the IPSR kernel estimate $R(t)$ versus $t$ (the band width of the Gaussian kernel estimate is 1 ms) in (a1)-(a5), and ISI histogram in (b1)-(b5) (each ISI histogram is composed of $5 \times 10^4$ ISIs, the bin size for the histogram is 0.5 ms, and vertical dotted lines in (b1)-(b5) denote integer multiples of the global period $T_G$ ($\simeq 6.8$ ms) of $R(t)$). Plots of (c1) the average occupation degree $\left<O_i\right>$, (c2) the average pacing degree $\left<P_i\right>$, and (c3) the statistical-mechanical spiking measure $M_s$ versus $p$.
}
\label{fig:Sync}
\end{figure}

We now study the population and individual behaviors of synchronized states for various values of $p > p_c^*$. Through comparison of the population behaviors with individual behaviors, one can understand fast sparsely synchronized states well. With increasing $p$ the zigzagness degree of partial stripes in the raster plots of spikes becomes reduced [see Figs.~\ref{fig:Sync}(a1)-\ref{fig:Sync}(a5)], and eventually for $p = p_{max}$ $(\sim 0.4)$, the raster plot becomes composed of vertical stripes without zigzag, and then the pacing degree between spikes becomes nearly the same. Hence, the amplitude of the IPSR kernel estimate $R(t)$ increases up to $p_{max}$, and then its value becomes saturated. For these values of $p$, $R(t)$ shows regular oscillation with the population frequency $f_p=147$ Hz, corresponding to the ultrafast rhythm (100-200 Hz). In contrast to population rhythm, individual neurons make stochastic and sparse discharges as Geiger counters. We collect $5 \times 10^4$ ISIs from all individual neurons and get the ISI histograms which are shown in Figs.~\ref{fig:Sync}(b1)-\ref{fig:Sync}(b5). Multiple peaks appear at multiples of the period $T_G$ $(=1/f_p \simeq 6.8$ ms) of $R(t)$. Hence, individual neurons exhibit stochastic phase locking leading to stochastic spike skipping (i.e., intermittent spikings phase-locked to $R(t)$ at random multiples of the period of $R(t)$). For these values of $p$, mean firing rates $f_i$ of individual neurons, corresponding to the inverse of the average ISI, is $33$ Hz, and hence each neuron makes an average firing very sparsely once during 4.5 population cycles. Consequently, for $p>p_c^*$ fast sparsely synchronized rhythms emerge.

By varying $p$ in the whole range of fast sparse synchronization, we also measure the degree of fast sparsely synchronized rhythms in terms of a realistic statistical-mechanical spiking measure $M_s$, based on the IPSR kernel estimate $R(t)$, which was developed in our recent work \cite{RM}. As shown in Figs.~\ref{fig:Sync}(a1)-\ref{fig:Sync}(a5), population spike synchronization may be well visualized in a raster plot of spikes. For a synchronized case, the raster plot is composed of partially-occupied stripes (indicating sparse synchronization). To measure the degree of the population synchronization seen in the raster plot, a statistical-mechanical spiking measure $M_s$, based on $R(t)$, was introduced by considering the occupation pattern and the pacing pattern of the spikes in the stripes \cite{RM}. The spiking measure $M_i$ of the $i$th stripe is defined by the product of the occupation degree $O_i$ of spikes (representing the density of the $i$th stripe) and the pacing degree $P_i$ of spikes (denoting the smearing of the $i$th stripe):
\begin{equation}
  M_i = O_i \cdot P_i.
\label{eq:SM}
\end{equation}
The occupation degree $O_i$ in the $i$th stripe is given by the fraction of spiking neurons:
\begin{equation}
   O_i = \frac {N_i^{(s)}} {N},
\end{equation}
where $N_i^{(s)}$ is the number of spiking neurons in the $i$th stripe. For sparse synchronization, $O_i<< 1$, while $O_i=1$ for full synchronization. The pacing degree $P_i$ of each microscopic spike in the $i$th stripe can be determined in a statistical-mechanical way by taking into account its contribution to the macroscopic IPSR kernel estimate $R(t)$. Each global cycle of $R(t)$ begins from its left minimum, passes the central maximum, and ends at the right minimum; the central maxima coincide with centers of stripes in the raster plot [see Figs.~\ref{fig:Sync}(a1)-\ref{fig:Sync}(a5)]. An instantaneous global phase $\Phi(t)$ of $R(t)$ is introduced via linear interpolation in the two successive subregions forming a global cycle \cite{RM,GP}. The global phase $\Phi(t)$ between the left minimum (corresponding to the beginning point of the $i$th global cycle) and the central maximum is given by
\begin{equation}
\Phi(t) = 2\pi(i-3/2) + \pi \left(
\frac{t-t_i^{(min)}}{t_i^{(max)}-t_i^{(min)}} \right)
 {\rm~~ for~} ~t_i^{(min)} \leq  t < t_i^{(max)}
~~(i=1,2,3,\dots),
\end{equation}
and $\Phi(t)$ between the central maximum and the right minimum (corresponding to the beginning point of the $(i+1)$th cycle) is given by
\begin{equation}
\Phi(t) = 2\pi(i-1) + \pi \left(
\frac{t-t_i^{(max)}}{t_{i+1}^{(min)}-t_i^{(max)}} \right)
 {\rm~~ for~} ~t_i^{(max)} \leq  t < t_{i+1}^{(min)}
~~(i=1,2,3,\dots),
\end{equation}
where $t_i^{(min)}$ is the beginning time of the $i$th global cycle (i.e., the time at which the left minimum of $R(t)$ appears in the $i$th global cycle) and $t_i^{(max)}$ is the time at which the maximum of $R(t)$ appears in the $i$th global cycle. Then, the contribution of the $k$th microscopic spike in the $i$th stripe occurring at the time $t_k^{(s)}$ to $R(t)$ is given by $\cos \Phi_k$, where $\Phi_k$ is the global phase at the $k$th spiking time [i.e., $\Phi_k \equiv \Phi(t_k^{(s)})$]. A microscopic spike makes the most constructive (in-phase) contribution to $R(t)$ when the corresponding global phase $\Phi_k$ is $2 \pi n$ ($n=0,1,2, \dots$) while it makes the most destructive (anti-phase) contribution to $R(t)$ when $\Phi_i$ is $2 \pi (n-1/2)$. By averaging the contributions of all microscopic spikes in the $i$th stripe to $R(t)$, we obtain the pacing degree of spikes in the $i$th stripe:
\begin{equation}
 P_i ={ \frac {1} {S_i}} \sum_{k=1}^{S_i} \cos \Phi_k,
\label{eq:PACING}
\end{equation}
where $S_i$ is the total number of microscopic spikes in the $i$th stripe.
By averaging $M_i$ of Eq.~(\ref{eq:SM}) over a sufficiently large number $N_s$ of stripes, we obtain the statistical-mechanical spiking measure $M_s$:
\begin{equation}
M_s =  {\frac {1} {N_s}} \sum_{i=1}^{N_s} M_i.
\label{eq:CM}
\end{equation}
By varying $p$, we follow $3 \times 10^3$ stripes and characterize sparse synchronization in terms of $\left<O_i\right>$ (average occupation degree), $\left<P_i\right>$ (average pacing degree), and the statistical-mechanical spiking measure $M_s$ for 12 values of $p$ in the sparsely synchronized region, and the results are shown in Figs.~\ref{fig:Sync}(c1)-\ref{fig:Sync}(c3). We note that the average occupation degree $\left<O_i\right>$ (denoting the average density of stripes in the raster plot) is nearly the same ($\left<O_i\right> \simeq 0.22$), independently of $p$; only a fraction (about 1/4.5) of the total neurons fire in each stripe [see Figs.~\ref{fig:Sync}(a1)-\ref{fig:Sync}(a5)]. This partial occupation in the stripes results from stochastic spike skipping of individual neurons and is seen well in the multi-peaked ISI histograms [see Figs.~\ref{fig:Sync}(b1)-\ref{fig:Sync}(b5)]. The average occupation degree ($\left<O_i\right> \simeq 0.22$) implies that individual neurons fire about once during the 4.5 global cycles, which agrees well with the average firing rates $(\simeq 33$ Hz) of individual neurons obtained from the ISI distributions shown in Figs.~\ref{fig:Sync}(b1)-\ref{fig:Sync}(b5). Hence, the average occupation degree $\left<O_i\right>$ characterize the sparseness degree of population synchronization well. On the other hand, with increasing $p$, the average pacing degree $\left< P_i\right>$ increases rapidly due to appearance of long-range connections. However, the value of $\left< P_i\right>$ saturates for $p = p_{max}$ $(\sim 0.4)$ because long-range short-cuts which appear up to $p_{max}$ play sufficient role to get maximal pacing degree, as in the case of pair cross-correlations in Figs.~\ref{fig:Corr}(e1)-\ref{fig:Corr}(e8). Figure \ref{fig:Sync}(c3) shows the statistical-mechanical spiking measure $M_s$ (taking into account both the occupation and the pacing degrees of spikes) versus $p$. As in the case of $\left< P_i\right>$, $M_s$ makes a rapid increase up to $p = p_{max}$, because $\left<O_i\right>$ is nearly independent of $p$. $M_s(p)$ is nearly equal to $\left< P_i\right>/4.5$ because of the sparse occupation [$\left<O_i\right> \simeq 1/4.5]$.

As the rewiring probability $p$ is increased from $p_c^*$, synchronization degree is increased because global efficiency of information transfer becomes better. However, with increasing $p$, the network axon wiring length becomes longer due to long-range short-cuts. Longer axonal connections are expensive because of material and energy costs. Hence, in view of dynamical efficiency we search for optimal population rhythm emerging at a minimal wiring cost. An optimal fast sparsely synchronized rhythm may emerge via tradeoff between the synchronization degree and the wiring cost.
The synchronization degree is given by the statistical-mechanical spiking measure $M_s$ shown in Fig.~\ref{fig:Sync}(c3). We then calculate the wiring length by varying $p$ on a ring of radius $r$ (=$N/ 2 \pi$) where neurons are placed equidistantly. The axonal wiring length, $L_{ij}$, between neuron $i$ and neuron $j$ is given by the arc length between two vertices $i$ and $j$ on the ring:
\begin{equation}
   L_{ij} = \left\{
\begin{array}{l}
   |j-i| \; \textrm{for} \; |j-i| \le \frac{N}{2}\\
   N-|j-i| \; \textrm{for} \; |j-i| > \frac{N}{2}.
\end{array}
\right.
\end{equation}
Then, the total wiring length is:
\begin{equation}
   L_{total} = \sum_{i=1}^{N} \sum_{j=1 (j \ne i)}^{N} a_{ij} \cdot L_{ij},
\end{equation}
where $a_{ij}$ is the $ij$ element of the adjacency matrix $A$ of the network. The connection between vertices in the network is represented by its $N \times N$ adjacency matrix $A$ $(=\{a_{ij} \})$ whose element values are $0$ or $1$. If $a_{ij}=1$, then an edge from the vertex $i$ to the vertex $j$ exists; otherwise no such edges exists. This adjacency matrix $A$ corresponds to the transpose of the connection weight matrix $W$ in Sec.~\ref{sec:SWN}. We get a normalized wiring length $\cal{L}$ by dividing $L_{total}$ with $L_{total}^{(global)}$ $[=\sum_{i=1}^{N} \sum_{j=1 (j \ne i)}^{N} L_{ij}]$ which is the total wiring length for the global-coupled case:
\begin{equation}
{\cal{L}}= \frac{L_{total}}{L_{total}^{(global)}}.
\end{equation}
\begin{figure}[t]
\includegraphics[width=0.8\columnwidth]{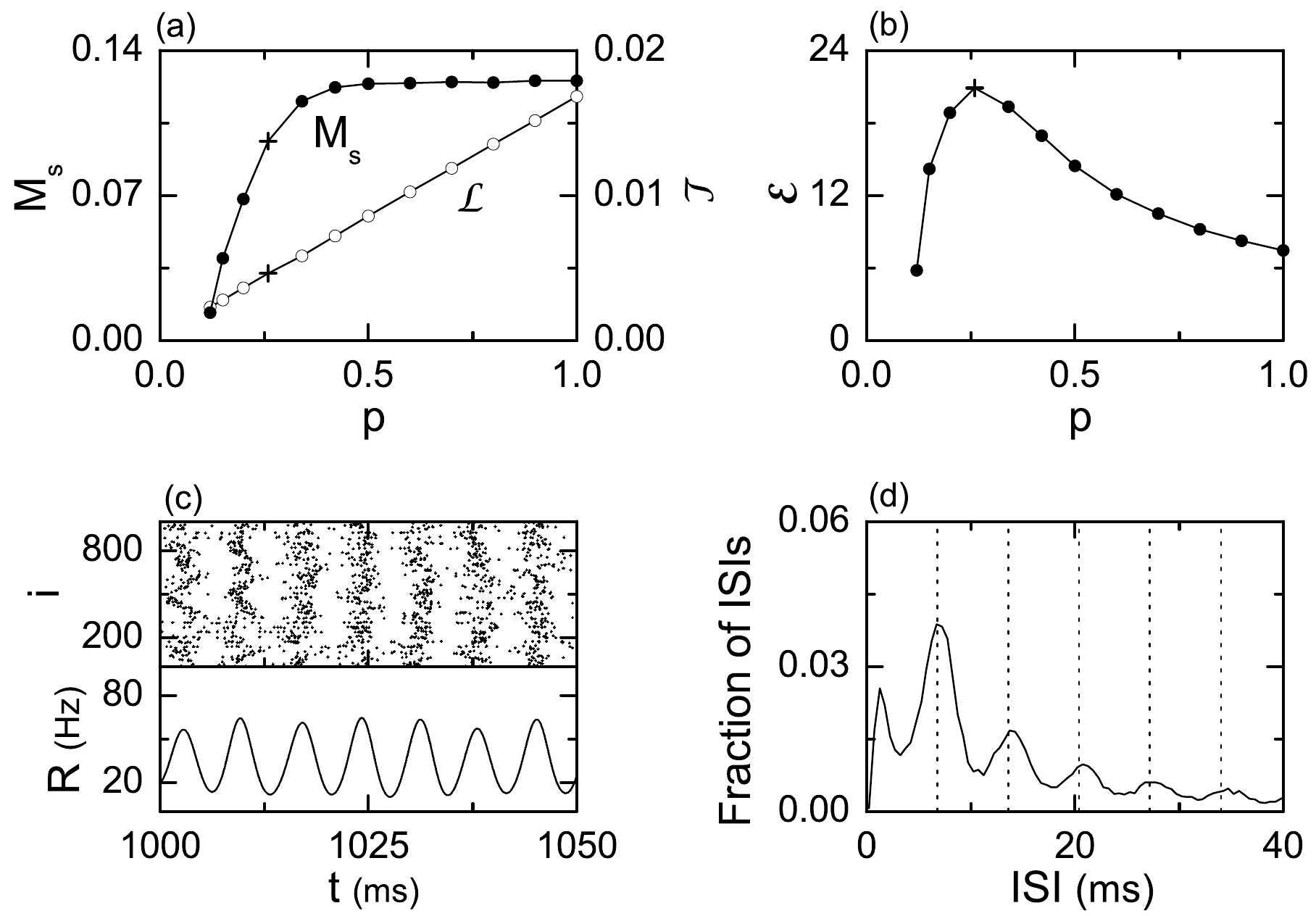}
\caption{Watts-Strogatz small-world network of suprathreshold FS Izhikevich interneurons for $N=10^3$, $I_{DC}=1500$, $M_{syn}=50$, $J=1400$, and $D=500$. (a) Statistical-mechanical spiking measure $M_s$ and normalized wiring length $\cal{L}$ versus $p$. (b) Dynamical efficiency $\cal{E}$ versus $p$. The values of $M_s$, $\cal{L}$, and $\cal{E}$ at an optimal value $p^*_{\cal{E}}$ $(\simeq 0.26)$ are denoted by the symbol ``$+$''. Optimally fast sparsely synchronized rhythm for $p=p^*_{\cal{E}}$: (c) raster plot of neural spikes and plot of the IPSR kernel estimate $R(t)$ versus $t$ and (d) ISI histogram (ISI histogram is composed of $5 \times 10^4$ ISIs, the bin size for the histogram is 0.5 ms, and vertical dotted lines in (d) denote integer multiples of the global period $T_G$ ($\simeq 6.8$ ms) of $R(t)$).
}
\label{fig:DE}
\end{figure}
Plot of $\cal{L}$ versus $p$ is shown in Fig.~\ref{fig:DE}(a). It increases linearly with respect to $p$. Hence, with increasing $p$, the wiring cost becomes expensive. An optimal rhythm may emerge through tradeoff between the synchronization degree $M_s$ and the wiring cost $\cal{L}$. To this end, a dynamical efficiency $\cal{E}$ is given by \cite{Buz2}:
\begin{equation}
 {\cal{E}}= \frac{\textrm{Synchrony\, Degree}~ (M_s)}{\textrm{Normalized\, Wiring\, Length}~ ({\cal{L}})}.
\end{equation}
Figure \ref{fig:DE}(b) shows plot of $\cal{E}$ versus $p$. For $p=p^*_{\cal{E}}$ $(\simeq 0.26)$, an optimal rhythm is found to emerge at a minimal wiring cost in an economic small-world network. An optimal fast sparsely synchronized rhythm is shown in Fig.~\ref{fig:DE}(c). Since the economical small-world network has a moderate clustering coefficient $C(p^*_{\cal{E}})$ $(=0.3)$, the raster plot of spikes shows a zigzag pattern due to local clustering of spikes, and the IPSR kernel estimate $R(t)$ exhibits a regular ultrafast oscillation at a population frequency $f_p$ $(=147$ Hz). In contrast to population rhythm, individual neurons fire irregularly and sparsely with $f_i=33$ Hz as Geiger counters, as shown well in the multi-peaked ISI histogram of Fig.~\ref{fig:DE}(d).

\section{Summary}
\label{sec:SUM}
We have investigated the effect of network architecture on fast sparsely synchronized cortical rhythms with stochastic and intermittent neural discharges. These fast sparsely synchronized neural oscillations are in contrast to fully synchronized oscillations with regular neural discharges. For modeling of complex connections in neural systems, we first used the conventional Erd\"{o}s-Renyi random graph of suprathreshold FS Izhikevich interneurons, and studied occurrence of the population synchronized states by varying the inhibition strength $J$ and the noise intensity $D$. Fast sparsely synchronized states have been found to appear for large values of $J$ and $D$. However, real synaptic connections are known to be neither regular nor random. Hence, we considered the Watts-Strogatz model for small-world networks which interpolates between the regular lattice and the random graph, and for fixed values of $J$ and $D$ ($J=1400$ and $D=500$), we investigated the effect of small-world connectivity on emergence of fast sparsely synchronized rhythms by varying the rewiring probability $p$ from local to long-range connections. Through calculation of a realistic thermodynamic order parameter $\cal{O}$, fast sparsely synchronized rhythms have been found to emerge as $p$ passes a small critical value $p_c^*$ $(\simeq 0.12)$. For $p > p_c^*$, the IPSR kernel estimate $R(t)$ has been found to oscillate with population frequency of 147Hz. However, individual neurons discharge spikes stochastically at low rates $(\sim 33$ Hz) which is much lower than the population frequency.
We have also investigated the effect of geometrical long-range connections on dynamical correlations between neuronal pairs for occurrence of global synchronization. It has thus been found that for $p > p_c^*$, the dynamical correlation length covers the whole system, thanks to sufficient number of long-range connections, and consequently global synchronization appears in the whole population. The degree of fast sparse synchronization has been well measured in terms of the realistic statistical-mechanical spiking measure $M_s$ introduced by considering both the occupation and the pacing degrees of spikes in the raster plot of neural spikes. As $p$ is increased, the synchrony degree increases, while the network axon wiring length also becomes longer because more long-range connections appear. Hence, wiring economy must be taken into account for dynamical efficiency. A ratio of the synchrony degree to the geometrical wiring cost is found to be maximal at a dynamical-efficiency optimal value $p^*_{\cal{E}}$ $(\simeq 0.26)$. For this case, an optimal fast sparsely synchronized rhythm is found to emerge at a minimal wiring cost in an economic small-world network.

\section*{Acknowledgments}
This research was supported by Basic Science Research Program through the National Research Foundation of Korea(NRF) funded by the Ministry of Education (Grant No. 2013057789).


\section*{References}
\begin{thebibliography}{}
\bibitem{Buz1} G. Buzs$\acute{\rm a}$ki, Rhythms of the Brain, Oxford University Press, New York, 2006.
\bibitem{W_Review} X.-J. Wang, Neurophysiological and computational principles of cortical rhythms in cognition, Physiological Reviews 90 (2010) 1195-1268.
\bibitem{TW} R.D. Traub, M.A. Whittington, Cortical Oscillations in Health and Diseases, Oxford University Press, New York, 2010.
\bibitem{SS1} E.H. Buhl, G. Tamas, A. Fisahn, Cholinergic activation and tonic excitation induce persistent gamma oscillations in mouse somatosensory cortex in vitro, Journal of Physiology 513 (1998) 117-126.
\bibitem{SS2} A. Fisahn, F.G. Pike, E.H. Buhl, O. Paulsen, Cholinergic induction of network oscillations at 40 Hz in the hippocampus in vitro, Nature 394 (1998) 186-189.
\bibitem{SS3} J. Csicsvari, H. Hirase, A. Czurko, G. Buzs$\acute{\rm a}$ki, Reliability and state dependence of pyramidal cell-interneuron synapses in the hippocampus: an ensemble approach in the behaving rat, Neuron 21 (1998) 179-189.
\bibitem{SS4} J. Csicsvari, H. Hirase, A. Czurko, A. Mamiya, G. Buzs$\acute{\rm a}$ki, Oscillatory coupling of hippocampal pyramidal cells and interneurons in the behaving rat, Journal of Neuroscience 19 (1999) 274-287.
\bibitem{SS5} J. Fellous, T.J. Sejnowski, Cholinergic induction of oscillations in the hippocampal slice in the slow (0.5-2 Hz), theta (5-12 Hz), and gamma (35-70 Hz) bands, Hippocampus 10 (2000) 187-197.
\bibitem{SS6} P. Fries, J.H. Reynolds, A.E. Rorie, R. Desimone, Modulation of oscillatory neuronal synchronization by selective visual attention, Science 291 (2001) 1560-1563.
\bibitem{SS7} N.K. Logothetis, J. Pauls, M.A. Augath, T. Trinath, A. Oeltermann, Neurophysiological investigation of the basis of the fMRI signal, Nature 412 (2001) 150-157.
\bibitem{WB} X.-J. Wang, G. Buzs$\acute{\rm a}$ki, Gamma oscillations by synaptic inhibition in a hippocampal interneuronal network, Journal of Neuroscience 16 (1996) 6402-6413.
\bibitem{Wang} X.-J. Wang, Neural oscillations, in: L. Nadel (Ed.), Encyclopedia of Cognitive Science, MacMillan, London, 2003, pp. 272-280.
\bibitem{Sparse1} N. Brunel, V. Hakim, Fast global oscillations in networks of integrate-and-fire neurons with low firing rates, Neural Computation 11 (1999) 1621-1671.
\bibitem{Sparse2} N. Brunel, Dynamics of sparsely connected networks of excitatory and inhibitory spiking neurons, Journal of Computational Neuroscience 8 (2000) 183-208.
\bibitem{Sparse3} N. Brunel, X.-J. Wang, What determines the frequency of fast network oscillations with irregular neural discharges? Journal of Neurophysiology 90 (2003) 415-430.
\bibitem{Sparse4} C. Geisler, N. Brunel, X.-J. Wang, The contribution of intrinsic membrane dynamics to fast network oscillations with irregular neuronal discharges, Journal of Neurophysiology 94 (2005) 4344-4361.
\bibitem{Sparse5} N. Brunel, D. Hansel, How noise affects the synchronization properties of recurrent networks of inhibitory neurons, Neural Computation 18 (2006) 1066-1110.
\bibitem{Sparse6} N. Brunel, V. Hakim, Sparsely synchronized neuronal oscillations, Chaos 18 (2008) 015113.
\bibitem{TJ1} P.H.E. Tiesinga, J.V. Jos\'{e}, Robust gamma oscillations in networks of inhibitory hippocampal interneurons, Network: Computation in Neural Systems 11 (2000) 1-23.
\bibitem{TJ2} P.H.E. Tiesinga, J.V. Jos\'{e}, Synchronous clusters in a noisy inhibitory neural network, Journal of Computational Neuroscience 9 (2000) 49-65.
\bibitem{Kim1} W. Lim, S.-Y. Kim, Statistical-mechanical measure of stochastic spiking coherence in a population of inhibitory subthreshold neuron, Journal of Computational Neuroscience
31 (2011) 667-677.
\bibitem{Kim2} D.-G. Hong, S.-Y. Kim, W. Lim, Effect of sparse random connectivity on the stochastic spiking coherence of inhibitory subthreshold neurons, Journal of the Korean Physical Society 59 (2011) 2840-2846.
\bibitem{Kim3} S.-Y. Kim, W. Lim, Sparsely-synchronized brain rhythm in a small-world neural network, Journal of the Korean Physical Society 63 (2013) 104-113.
\bibitem{Izhi1} E.M. Izhikevich, Simple model of spiking neurons, IEEE Transactions on Neural Networks 14 (2003) 1569-1572.
\bibitem{Izhi2} E.M. Izhikevich, Which model to use for cortical spiking neurons? IEEE Transactions on Neural Networks 15 (2004) 1063-1070.
\bibitem{Izhi3} E.M. Izhikevich, Dynamical Systems in Neuroscience, MIT Press, Cambridge, 2007.
\bibitem{Izhi4} E.M. Izhikevich, Hybrid spiking models, Philosophical Transactions of the Royal Society A 368 (2010) 5061-5070.
\bibitem{ER} P. Erd\"{o}s, A. Renyi, On random graphs I, Publicationes Mathematicae Debrecen 6 (1959) 290-297.
\bibitem{Eff1} V. Latora, M. Marchiori, Efficient behavior of small-world networks, Physical Review Letters 87 (2001) 198701.
\bibitem{Eff2} V. Latora, M. Marchiori, Economic small-world behavior in weighted networks, The European Physical Journal B 32 (2003) 249-263.
\bibitem{Sporns} O. Sporns, Networks of the Brain, MIT Press, Cambridge, 2011.
\bibitem{Buz2} G. Buzs$\acute{\rm a}$ki, C. Geisler, D.A. Henze, X.-J. Wang, Interneuron diversity series: circuit complexity and axon wiring economy of cortical interneurons, Trends in Neurosciences 27 (2004) 186-193.
\bibitem{CN1} D.B. Chklovskii, B.W. Mel, K. Svoboda, Cortical rewiring and information storage, Nature 431 (2004) 782-788.
\bibitem{CN2} S. Song, P.J. Sj$\ddot{\rm o}$str$\ddot{\rm o}$m, M. Reigl, S. Nelson, D.B. Chklovskii, Highly nonrandom features of synaptic connectivity in local cortical circuits, PLoS Biology 3 (2005) e68.
\bibitem{CN3} O. Sporns, C.J. Honey, Small worlds inside big brains, Proceedings of the National Academy of Sciences of the United States of America 103 (2006) 19219-19220.
\bibitem{CN4} P. Larimer, B.W. Strowbridge, Nonrandom local circuits in the dentate gyrus, Journal of Neuroscience 28 (2008) 12212-12223.
\bibitem{CN5} E. Bullmore, O. Sporns, Complex brain networks: Graph-theoretical analysis of structural and functional systems, Nature Reviews Neuroscience 10 (2009) 186-198.
\bibitem{CN6} O. Sporns, G. Tononi, G.M. Edelman, Theoretical neuroanatomy: Relating anatomical and functional connectivity in graphs and cortical connection matrices, Cerebral Cortex
10 (2000) 127-141.
\bibitem{CN7} D.S. Bassett, E. Bullmore, Small-world brain networks, The Neuroscientist 12 (2006) 512-523.
\bibitem{SWN1} D.J. Watts, S.H. Strogatz, Collective dynamics of `small-world' networks, Nature 393 (1998) 440-442.
\bibitem{SWN2} S.H. Strogatz, Exploring complex networks, Nature 410 (2001) 268-276.
\bibitem{SWN3} D.J. Watts, Small Worlds: The Dynamics of Networks Between Order and Randomness, Princeton University Press, 2003.
\bibitem{SDS1} S. Milgram, The small-world problem, Psychology Today 1 (1967) 61-67.
\bibitem{SDS2} J. Guare, Six Degrees of Separation: A Play, Random House, New York, 1990.
\bibitem{SW2} L.F Lago-Fern$\acute{\rm a}$ndez, R. Huerta, F. Corbacho, J.A. Sig$\ddot{\rm u}$enza, Fast response and temporal coherent oscillations in small-world networks, Physical Review Letters 84 (2000) 2758-2761.
\bibitem{SW3} O. Kwon, H.T. Moon, Coherence resonance in small-world networks of excitable cells, Physics Letters A 298 (2002) 319-324.
\bibitem{SW4} A. Roxin, H. Riecke, S.A. Solla, Self-sustained activity in a small-world network of excitable neurons, Physical Review Letters 92 (2004) 198101.
\bibitem{SW5} M. Kaiser, C.C. Hilgetag, Nonoptimal component placement, but short processing paths, due to long-distance projections in neural systems, PLoS Computational Biology 2 (2006) e95.
\bibitem{SW6} H. Riecke, A. Roxin, S. Madruga, S. Solla, Multiple attractors, long chaotic transients, and failure in small-world networks of excitable neurons, Chaos 17 (2007) 026110.
\bibitem{SW7} S. Achard, E.T. Bullmore, Efficiency and cost of economical brain functional networks, PLoS Computational Biology 3 (2007) e17.
\bibitem{SW8} S. Yu, D. Huang, W. Singer, D. Nikolie, A small world of neuronal synchrony, Cerebral Cortex 18 (2008) 2891-2901.
\bibitem{SW9} Q. Wang, Z. Duan, M. Perc, G. Chen, Synchronization transitions on small-world neuronal networks: Effects of information transmission delay and rewiring probability, Europhysics Letters 83 (2008) 50008.
\bibitem{SW10} M. Shanahan, Dynamical complexity in small-world networks of spiking neurons, Physical Review E 78 (2008) 041924.
\bibitem{SW11} M. Ozer, M. Perc, M. Uzuntarla, Stochastic resonance on Newman-Watts networks of Hodgkin-Huxley neurons with local periodic driving, Physics Letters A 373 (2009) 964-968.
\bibitem{SW12} Q. Wang, M. Perc, Z. Duan, G. Chen, Impact of delays and rewiring on the dynamics of small-world neuronal networks with two types of coupling, Physica A 389 (2010) 3299-3306.
\bibitem{SW13} J.T. Lizier, S. Pritam, M. Prokopenko, Information dynamics in small-world boolean networks, Artificial Life 17 (2011) 293-314.
\bibitem{Wiring1} S.B. Laughlin, T.J. Sejnowski, Communication in neuronal networks, Science 301 (2003) 1870-1874.
\bibitem{Wiring2} D.B. Chklovskii, T. Schikorski, C.F. Stevens, Wiring optimization in cortical circuits, Neuron 34 (2002) 341-347.
\bibitem{Wiring3} D.B. Chklovskii, A.A. Koulakov, Maps in the brain: what can we learn from them? Annual Review of Neuroscience 27 (2004) 369-392.
\bibitem{Wiring4} D.B. Chklovskii, Synaptic connectivity and neuronal morphology: two sides of the same coin, Neuron 43 (2004) 609-617.
\bibitem{Wiring6} O. Sporns, The non-random brain: efficiency, economy, and complex dynamics, Frontiers in Computational Neuroscience 5 (2011) 5.
\bibitem{Wiring7} E. Bullmore, O. Sporns, The economy of brain network organization, Nature Reviews Neuroscience 13 (2012) 336-349.
\bibitem{RM} S.-Y. Kim, W. Lim, Realistic thermodynamic and statistical-mechanical measures for neural synchronization, Journal of Neuroscience Methods 226 (2014) 161-170.
\bibitem{Ex1} A.L. Hodgkin, The local electric changes associated with repetitive action in a nonmedullated axon, Journal of Physiology 107 (1948) 165-181.
\bibitem{Ex2} E.M. Izhikevich, Neural excitability, spiking and bursting, International Journal of Bifurcation and Chaos 10 (2000) 1171-1266.
\bibitem{SDE} M. San Miguel, R. Toral, Stochastic effects in physical systems, in: J. Martinez, R. Tiemann, E. Tirapegui (Eds.), Instabilities and Nonequilibrium Structures VI, Kluwer Academic Publisher, Dordrecht, 2000, pp. 35-130.
\bibitem{Kernel} H. Shimazaki, S. Shinomoto, Kernel bandwidth optimization in spike rate estimation, Journal of Computational Neuroscience 29 (2010) 171-182.
\bibitem{GR} D. Golomb, J. Rinzel, Clustering in globally coupled inhibitory neurons, Physica D 72 (1994) 259-282.
\bibitem{Longtin1} A. Longtin, Synchronization of the stochastic Fitzhugh-Nagumo equations to periodic forcing, Nuovo Cimento D 17 (1995) 835-846.
\bibitem{Longtin2} A. Longtin, Stochastic aspects of neural phase locking to periodic signals, in: S. Kim, K. J. Lee, W. Sung (Eds.), Stochastic Dynamics and Pattern Formation in Biological and Complex Systems, AIP, New York, 2000, pp. 219-239.
\bibitem{Order1} D. Hansel, G. Mato, Asynchronous states and the emergence of synchrony in large networks of interacting excitatory and inhibitory neurons, Neural Computation 15 (2003) 1-56.
\bibitem{Order2} D. Hansel, H. Sompolinsky, Synchronization and computation in a chaotic neural network, Physical Review Letters 68 (1992) 718721.
\bibitem{Order3} I. Ginzburg, H. Sompolinsky, Theory of correlations in stochastic neural networks, Physical Review E 50 (1994) 3171-3191.
\bibitem{GP} J. Freund, L. Schimansky-Geier, P. H$\ddot{\rm a}$nggi, Frequency and phase synchronization in stochastic systems, Chaos 13 (2003) 225-238.
\end{thebibliography}
\end{document}